\documentclass[a4paper]{article}

\usepackage[margin=1in]{geometry} 

\usepackage{amsmath}
\usepackage{amsthm}
\usepackage{amssymb}
\usepackage{esint}

\usepackage[utf8]{inputenc}
\usepackage[hidelinks]{hyperref}
\hypersetup{
	unicode,
	pdfauthor={Author One, Author Two, Author Three},
	pdftitle={OPY2025},
	pdfsubject={A simple article template},
	pdfkeywords={article, template, simple},
	pdfproducer={LaTeX},
	pdfcreator={pdflatex}
}

\usepackage[sort&compress,numbers,square]{natbib}
\theoremstyle{plain}
\newtheorem{theorem}{Theorem}

\theoremstyle{definition}

\newtheorem{remark}[theorem]{Remark}

\def\@#1{{\mathbf{#1}}}

\def\Xint#1{\mathchoice
{\XXint\displaystyle\textstyle{#1}}%
{\XXint\textstyle\scriptstyle{#1}}%
{\XXint\scriptstyle\scriptscriptstyle{#1}}%
{\XXint\scriptscriptstyle\scriptscriptstyle{#1}}%
\!\int}
\def\XXint#1#2#3{{\setbox0=\hbox{$#1{#2#3}{\int}$ }
\vcenter{\hbox{$#2#3$ }}\kern-.575\wd0}}

\def\dashint{\Xint-}

\DeclareMathOperator{\sech}{sech}

\usepackage{graphicx, color}
\graphicspath{{fig/}}

\usepackage{comment}
\usepackage[dvipsnames,svgnames]{xcolor}

\def\gl{\mathrel{\mathpalette\overl@ss>}}

\def\sech{\mathop{\rm sech}\nolimits}

\def\Real{\mathbb{R}}

\def\Complex{\mathbb{C}}

\def\i{\text{i}}

\def\Re{\mathop{\rm Re}\nolimits}
\def\Im{\mathop{\rm Im}\nolimits}
\def\Res{\mathop{\rm Res}\limits}

\def\sgn{\mathop{\rm sgn}\nolimits}
\def\e{\mathop{\rm e}\nolimits}
\def\@#1{{\mathbf{#1}}}
\def\_#1{{\mathsf{#1}}}
\let\~=\tilde
\let\==\bar
\let\^=\hat
\let\<=\langle
\let\>=\rangle

\def\be{\begin{equation}}
	\def\ee{\end{equation}}
\def\bse{\begin{subequations}}
	\def\ese{\end{subequations}}
\usepackage{algorithm, algpseudocode} 
\usepackage{mathrsfs} 

\usepackage{lipsum}
\numberwithin{equation}{section}
\title{Integrable perturbation theory for dark solitons of \\the defocusing nonlinear Schr\"odinger equation}
\author{Nicholas J. Ossi$^1$,\;\; Barbara Prinari$^{1,2,\dagger}$,\;\; Jianke Yang$^3$}
\date{
    \small{$^1$ Department of Mathematics, State University of New York, Buffalo, NY 14260, USA\\
    $^2$ Department of Mathematics, University of Ioannina, Ioannina 45110, Epirus, Greece\\
    $^3$ Department of Mathematics and Statistics, University of Vermont, Burlington, VT 05401, USA}\\
    $^\dagger$ \href{mailto:bprinari@buffalo.edu}{bprinari@buffalo.edu}
}
\begin{document}
	\maketitle
	\begin{abstract}
The goal of this work is to revisit the eigenfunction-expansion-based perturbation theory of the defocusing nonlinear Schr\"odinger equation on a nonzero background, and develop it to correctly predict the slow-time evolution of the dark soliton parameters, as well as the radiation shelf emerging on the soliton sides. Proof of the closure of the squared eigenfunctions is provided, and the complete set of eigenfunctions of the linearization operator is used to expand the first-order perturbation solution. Our closure/completeness relation accounts for the singularities of the scattering data at the branch points of the continuous spectrum, which leads to the correct discrete eigenfunctions. Using the one-soliton closure relation and its correct discrete eigenmodes, the slow-time evolution equations of the soliton parameters are determined. Moreover, the first-order correction integral to the dark soliton is shown to contain a pole due to singularities of the scattering data at the branch points. Analysis of this integral leads to predictions for the shelves, as well as a formula for the slow time evolution of the soliton’s phase, which in turn allows one to determine the slow-time dependence of the soliton center. All the results are corroborated by direct numerical simulations, and compared with earlier results.\\\\
	\end{abstract}

	\tableofcontents
	
\section{Introduction}
\label{S1}
The mathematical modeling of physical phenomena often leads to a certain class of nonlinear partial differential equations (PDEs) known as integrable systems. Distinguished features of integrable systems are that: (i) they admit soliton solutions, and (ii) their initial-value problem can be effectively linearized via the Inverse Scattering Transform (IST). One of the prototypical integrable systems is the nonlinear Schr\"odinger (NLS) equation:
\begin{equation*}
iq_t+q_{xx}-2\sigma |q|^2q=0, \qquad \sigma=\pm 1,
\end{equation*}
where $q(x,t)$ is a complex function of $x,t\in \Real$, subscripts denote partial differentiation throughout, and $\sigma$ distinguishes between anomalous and normal dispersion, with $\sigma=-1$ corresponding to the ``focusing'' NLS, and $\sigma=1$ to the ``defocusing'' NLS.
The NLS equations appear as universal models for weakly dispersive nonlinear wave trains, and have been derived in such diverse fields as deep water waves, plasma physics, nonlinear optical fibers, low-temperature physics and Bose-Einstein condensates, magneto-static spin waves, and more  \cite{Zak68,Benney69,Zak72,Hasegawa73a,Hasegawa73b,Kali77,Zvezdin83,Pet02}.

Unlike its focusing counterpart, the defocusing NLS does not admit localized bright solitons, exponentially decaying as $|x|\to \infty$, but it possesses dark soliton solutions, which appear as localized dips of intensity over a nonzero background \cite{KLD98,Panosbook}. We consider the defocusing NLS equation in the form:
\begin{equation}
\label{e:NLS}
    iq_{t}+q_{xx}-2(|q|^{2}-q_{0}^{2})q=0,
\end{equation}
with the constant nonzero boundary conditions
\begin{equation}
    \label{e:BC}
    q(x,t)\rightarrow q_{\pm}\equiv q_{0}e^{i\theta_\pm} \qquad \text{as } x\rightarrow\pm\infty,
\end{equation}
where $q_{0}>0$ is the (symmetric) amplitude of the background and $\theta_\pm\in \Real$ are the asymptotic phases; the additional linear term proportional to $q$ in \eqref{e:NLS} is introduced via a gauge transformation $q(x,t)\mapsto e^{-2iq_0^2t}q(x,t)$ in order to allow for time-independent boundary conditions. The exact single dark soliton solution of \eqref{e:NLS} is given by:
\begin{equation}
\label{e:gen_dark_sol}
q(x,t)=
e^{i\sigma_1}\left\{k_{1}+i\Lambda_{1}\tanh\left[\Lambda_{1}(x+2k_{1}t-x_{1})\right]\right\},
\end{equation}
where 
$-q_0<k_1<q_0$ determines the soliton velocity, while $\Lambda_1$ fixes its amplitude (i.e., the depth of the soliton below the background $q_0$) and $x_1,\sigma_1\in \Real$ give the soliton center and phase, respectively. Note that the dark soliton amplitude and velocity are not independent, since they are related to each other and to the background amplitude $q_0$ via: 
\begin{equation}
k_1^2+\Lambda_1^2=q_0^2.
\end{equation}
In addition, the phases $\theta_\pm$ of the background are related to the soliton parameters via:
\begin{equation}
 \theta_\pm =\sigma_1\pm \arctan (\Lambda_1/k_1).   
\end{equation}
One limitation of integrable models is that, in general, the systems considered in physical experiments are non-integrable. On the other hand, the theoretical predictions for the soliton solutions in integrable cases provide an extremely valuable tool for investigating non-integrable solitary waves in regimes that are not too far from the integrable ones.
As such, researchers rely on perturbation-based techniques of related integrable systems, when possible, to study how solitons and their evolution are affected by the inclusion in the mathematical description of terms that account for small dissipation, small linear/nonlinear loss, etc.

The perturbation theory for solitons that decay rapidly at infinity has been extensively explored since the late 1970s — using various different methods such as multi-scale perturbation analysis, IST-based techniques, eigenfunction-expansion techniques, perturbations of conserved quantities, and direct numerical simulations \cite{Kaup1976,Kaup1978,Kaup1990,Elgin,KarpmanZETF1977,KodamaMJASAM1980,Kivshar_rev1989,HermanJPA1990,JYangPRE1999,Yang2000}. On the other hand, the nonvanishing background characteristic of dark solitons introduces significant difficulties when one attempts to apply these perturbative approaches developed for the rapidly decaying case.

As mentioned above, for the scalar defocusing NLS equation, dark solitons are completely determined by the four parameters: $q_0$, $\Lambda_1$ (or $k_1)$, $x_1$, $\sigma_1$, which in general under perturbation develop an
adiabatic evolution (e.g., they acquire a non-trivial dependence on a slow time variable $T=\varepsilon t$). 
Some early works investigated the perturbation of black (i.e., stationary dark) solitons in lossy fibers numerically \cite{ZhaoOL1989} and later on analytically \cite{GianniniJQE1990,LisakOL1991}. In \cite{GianniniJQE1990}, a direct perturbation theory was developed based on a formal series expansion of the solution where the $x$ and $\varepsilon t$ dependencies are separated out, and the first two terms in the expansion are computed by direct integration. The method developed in \cite{LisakOL1991}, based on perturbed conserved quantities, was subsequently extended to gray (i.e., non-stationary dark) solitons and to generic perturbations, but only two of the four main soliton parameters, $q_0$ and $\Lambda_1$, were determined.
In \cite{KivsharPRE1994}, it was demonstrated that, under perturbation, the background evolves independently of the soliton. By separating the background amplitude from the soliton ``core'', the authors were able to determine the soliton’s amplitude and width through a Hamiltonian method based on perturbed conservation laws. Other works investigated instability-induced dynamics of dark solitons and oscillations of dark solitons in trapped  Bose-Einstein condensates \cite{Peli1996,PeliFraKev2005,Kivshar2014}.
It is important to note that, for dark solitons, the adiabatic evolution of the soliton parameters alone does not fully characterize the perturbed solution. This is because the perturbation produces a moving shelf on both sides of the soliton. The presence of this shelf —confirmed both numerically and analytically— was in fact used in \cite{BCJosaB1997} to account for discrepancies observed in the perturbed conservation laws, although the soliton core parameters were not determined analytically.
Shelves emerging from the wake of a bright soliton had already been observed in soliton perturbation theories for the Korteweg-de Vries (KdV) equation  \cite{KN1980,AS1981}, the fifth-order KdV equation \cite{Yang2001}, and the complex modified KdV \cite{Yang2003,Yang}. For this class of problems, it was possible to determine the bright soliton parameters and the shelf parameters by either direct perturbation theory \cite{KodamaMJASAM1980} or by using the squared eigenfunctions, as  detailed in \cite{Yang}.

To date, the most comprehensive analysis of dark-soliton perturbations for the scalar defocusing NLS is that of Ablowitz \textit{et al.} in \cite{FrantzPRSA2011}. Using a multi-scale expansion together with perturbed conservation laws, the authors derived both the magnitude and phase of the shelf, as well as the adiabatic evolution of all soliton parameters, showing the emergence of a moving boundary layer that links the inner soliton core to the outer background. This approach was recently generalized to describe the effect of small perturbation to the dark-bright solitons of the coupled NLS equation, the so-called Manakov system \cite{CFHKP2025} (see also \cite{Rothos2024}, where the perturbed evolution of the soliton parameters and the background were determined from variations of the Riemann-Hilbert problem (RHP)).  

An alternative approach to soliton perturbation theory in the rapidly decaying case was introduced in \cite{Kaup1976,Kaup1978,Elgin}. In this method, based on the IST, one calculates the variations of the scattering data of the soliton due to the perturbation, and then uses inverse scattering to reconstruct the perturbed solution. In the process, squared eigenfunctions (i.e., quadratic combinations of Jost eigenfunctions and their adjoints) play a critical role. 
Another soliton perturbation theory that solves the first-order perturbation equation directly was also developed in \cite{Fogel1977, Keener1977, Kaup1990} (see the monograph \cite{Yang} and references therein). In this method, the first-order perturbation equation is solved by expanding its solution into a set of complete eigenfunctions of the linearization operator. This method does not explicitly use the IST, and it is often easier to apply. But its connection to the IST is still critical, since the eigenfunctions of the linearization operator are simply the squared eigenfunctions in the IST. We will refer to this latter perturbative approach interchangeably as ``integrable'' perturbation theory or eigenfunction-expansion-based perturbation theory. 

Since the early 1990s, numerous efforts have been made to extend the integrable perturbation theory to dark solitons. In 1994, Konotop and Vekslerchik derived orthogonality conditions from a set of squared eigenfunctions for the scalar defocusing NLS equation on a constant background, from which all soliton parameters can, in principle, be obtained \cite{KonotopPRE1994}. However, this early work did not take into account the background evolution induced by the perturbation. 
Subsequent attempts at rigorous proofs of the completeness of the squared eigenfunctions followed: in \cite{ChenJPA1997} Chen, Chen \& Huang showed completeness of the squared eigenfunctions in the case of one-soliton,
and used it to investigate analytically a linear damping perturbation, and in \cite{ChenCPL1998} applied the method again to a self-steepening perturbation. 
In \cite{HuangJPA1999}, Huang, Chi \& Chen used a generalized Marchenko equation and extended the completeness proof of the previous work \cite{ChenJPA1997} to the multi-soliton case. The proof in \cite{ChenJPA1997} was then claimed to be incorrect by Ao \& Yan in \cite{AoYanJPA2005,AoJPA2006}, based on the observation that the complete set should have two, not just one, continuous spectrum basis vector, which resulted in different predictions for the soliton velocity and the first-order correction. 
In a subsequent work by Ao \& Yan \cite{AoYanCTP2007}, the results of \cite{ChenJPA1997,HuangJPA1999} and \cite{AoYanJPA2005,AoJPA2006} were then declared to be `equivalent' under an appropriate ``transformation between two integral variables''. As a matter of fact, the difference between the two can be traced to the fact that the earlier work used one of the symmetries of the scattering problem to reduce the number of eigenfunctions involved in the closure relation to only the ones that are linearly independent.
It is also worth mentioning that in \cite{LashkinPRE2004} squared eigenfunctions were used (though without explicitly referring to them, or to their completeness) to develop an eigenfunction-expansion-based perturbation theory for the defocusing NLS on a background, with no reference to the work in \cite{ChenJPA1997,HuangJPA1999} (conversely, the papers \cite{AoYanJPA2005,AoJPA2006,AoYanCTP2007} do not reference \cite{LashkinPRE2004}). 
An integrable perturbation theory for the dark-bright solitons of the defocusing Manakov equation 
was also developed in \cite{Rothos2016}.

Importantly, in all the earlier implementations of the dark-soliton perturbation theory that is 
based on eigenfunction expansions \cite{KonotopPRE1994,ChenJPA1997,ChenCPL1998,HuangJPA1999,AoYanJPA2005,AoJPA2006,AoYanCTP2007,LashkinPRE2004,Rothos2016},
some of the orthogonality conditions were flawed, because the discrete eigenmodes that were used did not account for the contributions from the poles at the branch points of the integral term in the completeness relation. As we will discuss in details in Sec~\ref{S5} and Sec~\ref{S7}, this resulted in erroneous evolution equations for (at least) some of the soliton parameters.
Moreover, none of the above references attempted to determine the radiation shelf that develops around the dark soliton, or presented comparisons of the theoretical predictions with numerical simulations. The existence of the shelf has long been confirmed by numerical simulations, and, as was shown in \cite{FrantzPRSA2011}, it is critical in developing the perturbation theory and contributes to the integrals used to determine the evolution of the soliton parameters. 
For instance, the adiabatic evolution of the soliton center in \cite{FrantzPRSA2011} is markedly different from the one obtained by expanding the solution in terms of squared eigenfunctions in \cite{KonotopPRE1994,ChenJPA1997,ChenCPL1998,HuangJPA1999,AoYanJPA2005,AoJPA2006,AoYanCTP2007,LashkinPRE2004,Rothos2024}, and it was conjectured in \cite{FrantzPRSA2011} that, because of the existence of the expanding shelf, the squared eigenfunctions associated with the soliton are an insufficient basis, questioning in fact the existence of a closure relation for this problem. 

The goal of this work is to revisit the perturbation theory of the scalar defocusing NLS on a nontrivial background based on the squared eigenfunctions, and develop it so that it can correctly predict the slow-time evolution of all the dark soliton parameters, as well as the radiation shelves emerging on the sides of the soliton. First, we prove completeness of the squared eigenfunctions for general potentials. A crucial difference with respect to the earlier works is the need to properly account for the singularities of the scattering data at the points $\pm q_0$, which are branch points for the continuous spectrum. Taking contributions from such singularities out of the integral term of the closure relation and leaving the remaining integral as a principal-value integral proves to be critical in this eigenfunction-expansion-based perturbation theory, because this leads to the correct discrete eigenmodes to be used in orthogonality conditions of the perturbation theory. Then we use the 1-soliton closure relation and suitable suppression of secular growth to determine the adiabatic evolution of the soliton parameters $k_1(T),\Lambda_1(T)$, as well as a condition relating the evolution of $\sigma_1(T)$ and $x_1(T)$ under perturbations of order $\varepsilon$ as functions of a slow time variable $T=\varepsilon t$. The latter condition is missing in all the previous works that used eigenfunction-expansion-based perturbation theories.
More importantly, the first-order correction integral to the dark soliton computed via squared eigenfunctions is shown to contain a pole due to singularities of the scattering data at the branch points. Analysis of this first-order correction integral leads to predictions for the height and velocity of the shelves. Moreover, this same analysis provides the spatial phase gradient on each side of the shelf, as well as a formula for the slow time evolution of the core soliton's phase, which in turn allows one to determine the dependence of the soliton center on the slow time.
All the results are corroborated by direct numerical simulations, and compared with the results of the direct perturbation theory in \cite{FrantzPRSA2011}, and with the earlier works using integrable perturbation theory. In particular, the estimates we obtain for all soliton parameters agree with the ones in \cite{FrantzPRSA2011} to order $\varepsilon$, the only difference being for the soliton center, which, unlike the other parameters, obtained from perturbed conserved quantities up to $\mathcal{O}(\varepsilon)$,  is determined in \cite{FrantzPRSA2011} in terms of differential equations obtained from the Hamiltonian at $\mathcal{O}(\varepsilon^2)$. We will also explain the shortcomings of the predictions for the adiabatic evolution of the soliton parameters in the earlier works within the framework of the eigenfunction-expansion-based perturbation theory.


The plan of the paper is the following. In Sec~\ref{S2} we give an overview of the IST in order to set the notations and present the properties of eigenfunctions and scattering data that are necessary for the following sections.
In Sec~\ref{S3} we derive the closure relation of squared eigenfunctions for general potentials by calculating the variations in the IST generalized to the case of a nonzero background. In Sec~\ref{S4} we discuss the linearization operator and obtain the 1-soliton closure relation. The multiple scale perturbation theory is developed in Sec~\ref{S5}. Sec~\ref{S6} investigates the first order correction, the shelf and evolution of the phase. The comparison with direct numerical simulations and earlier results is provided in Sec~\ref{S7}, with explicit applications to linear and nonlinear damping, dissipation and self-steepening. Finally, Sec~\ref{S8} is devoted to some concluding remarks and open problems, and more technical calculations are collected in the appendices.

\section{Overview of the IST}
\label{S2}
In this section we give a succinct overview of the IST for the defocusing NLS with nonzero boundary conditions, to set the notations and review the properties of eigenfunctions and scattering data which are required for the following sections. Additional details can be found, for instance, in \cite{FT1987,P2023}.

We start by considering the (unperturbed) defocusing NLS in the form
\eqref{e:NLS} with the constant symmetric nonzero boundary conditions
\begin{equation}
\label{e:BC_pm}
    q(x,t)\rightarrow q_{\pm}\equiv q_{0}e^{\pm i\theta}\qquad \text{as }x\rightarrow\pm\infty, \qquad q_0>0,\ \theta\in \Real.
\end{equation}
The asymptotic phases can be chosen as $\pm \theta$ without loss of generality on account of the invariance of the PDE upon multiplication by an arbitrary phase. [Note that arbitrary asymptotic phases $\theta_\pm$ as $x\to \pm \infty$ can be simply accounted for by replacing $\theta=(\theta_+-\theta_-)/2$ throughout.]
As shown in \cite{Gallo2004,Gallo2008,DPVV2013,CJ2016}, the IST can be  formulated as a bijection for an initial condition such that $q(x,0)-\tilde{q}(x)\in H^{1,1}(\Real)$, where 
$$
\tilde{q}(x)=q_o\left[ \cos \theta +i \sin \theta \tanh x \right],    
$$
\begin{equation*}
H^\ell(\Real):=\left\{ 
f:\Real \to \Complex \ \text{s.t.}\ 
f^{(k)}\in L^2(\Real), \  k=0,\dots,\ell
\right\}, \quad
H^{1,\ell}(\Real):=L^{2,\ell}(\Real)\cap H^\ell(\Real),
\end{equation*}
$L^p(\Real)$ are the standard Lebesgue spaces, and the weighted spaces $L^{p,s}(\Real)$ with norm defined as
\begin{equation*}
||f||_{L^{p,s}(\Real)}:=\left(
\int_\Real \langle x \rangle^{2s}|f(x)|^p\, dx \right)^{1/p},
\qquad
\langle x\rangle:=\sqrt{1+x^2}.
\end{equation*}
Additional smoothness and decay of the reflection coefficient are established in \cite{CJ2016} under stronger assumptions on the initial condition (e.g., by considering $H^{1,\ell}(\Real)$ for $\ell=3/2,2$, see also \cite{GPT2026} for a review).

The Lax pair of the defocusing NLS is given by
\bse
\label{e:Lax}
\begin{eqnarray}
\label{e:lax_x}
    &v_{x}=Uv,\qquad &U=-ik\sigma_{3}+Q,\qquad \sigma_{3}=\begin{bmatrix}1&0\\0&-1\end{bmatrix},\qquad Q=\begin{bmatrix}0&q\\q^{*}&0\end{bmatrix},\\
\label{e:lax_t}
    &v_{t}=Vv,\qquad
    &V=i\sigma_{3}(-2k^{2}+Q_{x}-Q^{2}+q_{0}^{2})+2kQ.
\end{eqnarray}
\ese
The asymptotic eigenvalues $\lambda$ of \eqref{e:lax_x} as $x\to \pm \infty$ are related to the spectral parameter $k$ by $\lambda^{2}=k^{2}-q_{0}^{2}$. Thus, $\lambda=\lambda(k)$ is a multivalued function with branch points at $k=\pm q_{0}$, and a two-sheeted Riemann surface cut along $(-\infty, -q_0)\cup (q_0,+\infty)$ can be defined such that $\Im \lambda\ge 0$ on one sheet, and $\Im \lambda\le 0$ on the other. A uniformization variable $z$ can be introduced in the following way
\begin{equation}
\label{e:uni}
    z=k+\lambda\,, \qquad k=\frac{1}{2}(z+q_{0}^{2}z^{-1})\,, \qquad \lambda=\frac{1}{2}(z-q_{0}^{2}z^{-1}),
\end{equation}
such that both sheets of the Riemann surface on which $k$ is defined are mapped into a single complex $z$-plane. The matrix Jost solutions that satisfy both parts of the Lax pair \eqref{e:Lax} are defined by the boundary conditions
\bse
\label{e:jost}
\begin{gather}
    \label{e:jost_phi}
    \Phi(x,t,z)=\begin{bmatrix}
        \phi(x,t,z)&\bar\phi(x,t,z)
    \end{bmatrix} \sim  E_{-}(z)e^{-i\Omega(x,t,z)\sigma_{3}} \qquad \text{as } x\rightarrow-\infty,\\
    \label{e:jost_psi}
    \Psi(x,t,z)=\begin{bmatrix}
        \bar\psi(x,t,z)&\psi(x,t,z)
    \end{bmatrix} \sim  E_{+}(z)e^{-i\Omega(x,t,z)\sigma_{3}}\qquad \text{as } x\rightarrow+\infty,
\end{gather}
\ese
where 
\begin{subequations}
\begin{gather}
\label{e:defEpm}
E_{\pm}(z)=I-iz^{-1}\sigma_{3}Q_{\pm}\equiv \begin{bmatrix} 1 & -iq_\pm/z \\iq^{*}_\pm/z & 1 \end{bmatrix},
        \quad
        Q_\pm =\begin{bmatrix}0&q_\pm \\q^{*}_\pm &0\end{bmatrix}, \\
\label{e:omega}
\Omega(x,t,z)=\lambda(z)(x+2k(z)t).
\end{gather}
\end{subequations} 
Here and in the following, $I$ denotes the $2\times 2$ identity matrix. It can be shown that if $q\to q_\pm$ sufficiently rapidly as $x\to \pm \infty$, the eigenfunctions $\phi$ and $\psi$ are analytic for $\Im z>0$ and continuous for  $\Im z\ge 0$, while $\bar\phi$ and $\bar\psi$ are analytic for $\Im z<0$ and continuous for $\Im z\le 0$, in both cases including at the images of the branch points $z=\pm q_0$. Furthermore, $\Phi$ and $\Psi$ have a singularity at $z=0$ (cf. \eqref{e:defEpm}), and
\begin{equation}
    \label{e:gamma_def}
    \det\Phi(x,t,z)=\det\Psi(x,t,z)=\gamma(z)\,,\qquad \gamma(z)=1-q_{0}^{2}z^{-2},
\end{equation}
which is nonzero except at $z=\pm q_{0}$. 
For $z\in\mathbb{R}\backslash\{\pm q_{0}\}$, the Jost solutions can be related via:
\begin{equation}
\label{e:scattering_matrix}
    \Phi(x,t,z)=\Psi(x,t,z)S(z)\,,\qquad S(z)=\begin{bmatrix}
        a(z)&\bar{b}(z)\\b(z)&\bar{a}(z)
    \end{bmatrix}\,,\qquad \det S(z)=1.
\end{equation}
The scattering coefficients $a(z)$ and $\bar{a}(z)$ can be analytically continued into the upper and lower half planes, respectively, while $b(z)$ and $\bar{b}(z)$ are only defined on the real axis. Importantly, since the eigenfunctions are defined as simultaneous solutions of the Lax pair, the scattering coefficients are time-independent. Specifically, they can be written as:
\bse
\label{e:wron}
\begin{eqnarray}
a(z)=\frac{W(\phi(x,t,z),\psi(x,t,z))}{\gamma(z)},\qquad \bar{a}(z)=\frac{W(\bar{\psi}(x,t,z),\bar{\phi}(x,t,z))}{\gamma(z)},\\
b(z)=\frac{W(\bar\psi(x,t,z),\phi(x,t,z))}{\gamma(z)},\qquad  \bar{b}(z)=\frac{W(\bar\phi(x,t,z),\psi(x,t,z))}{\gamma(z)},
\end{eqnarray}
\ese
where $W$ denotes the Wronskian determinant, showing that generically all scattering coefficients are singular as $z\to \pm q_0$, unless the Jost functions become linearly dependent at either $\pm q_{0}$, in which case $\pm q_{0}$ is called a ``virtual level'', and the scattering coefficients are $\mathcal O(1)$ as $z$ approaches $\pm q_0$. Importantly, for any reflectionless/pure soliton potential, both $\pm q_0$ are virtual levels, and all scattering coefficients are finite as $z\to \pm q_0$.

The eigenfunctions satisfy the following symmetries corresponding to the involution $z\mapsto z^{*}$:
\begin{equation}
\label{e:phi_psi_sym_1}
    \Phi(x,t,z)=\sigma_{1}\Phi^{*}(x,t,z^{*})\sigma_{1},\qquad
    \Psi(x,t,z)=\sigma_{1}\Psi^{*}(x,t,z^{*})\sigma_{1},
\end{equation}
which in turn imply the following symmetries for the scattering coefficients:
\begin{equation}
\label{e:ab_sym_1}
    a(z)=\bar{a}^{*}(z^{*})\quad \text{for } \Im z\geq 0, \qquad 
    b(z)=\bar{b}^{*}(z) \quad \text{for } z\in\mathbb{R}.
\end{equation}
Due to a second involution in the Lax pair, namely $z\mapsto q_{0}^{2}/z$, we also have the symmetries:
\begin{equation}
\label{e:phi_psi_sym_2}
    \Phi(x,t,z)=-\frac{1}{iz}\Phi(x,t,q_{0}^{2}/z)\sigma_{3}Q_{-}\, ,\qquad
    \Psi(x,t,z)=\frac{1}{iz}\Psi(x,t,q_{0}^{2}/z)\sigma_{3}Q_{+}\, ,
\end{equation}
and the scattering coefficients satisfy:
\begin{equation}
\label{e:ab_sym_2}
    a(q_0^2/z)=e^{2i\theta}\bar{a}(z)\quad \text{for } \Im z\geq 0, \qquad 
    b(q_0^2/z)=-\bar{b}(z)\quad \text{for } z\in\mathbb{R}.
\end{equation}
Eq.~\eqref{e:scattering_matrix} can be written as:
\begin{gather}
\label{e:rh}
\frac{\phi(x,t,z)}{a(z)}=\bar{\phi}(x,t,z)+\rho(z)\psi(x,t,z),\qquad
\frac{\bar \phi(x,t,z)}{\bar a(z)}={\psi}(x,t,z)+\bar\rho(z)\bar \psi(x,t,z),
\end{gather}
where the reflection coefficients are give by:
\begin{equation}
    \label{e:ref_def}
    \rho(z)=\frac{b(z)}{a(z)},\qquad \bar\rho(z)=\frac{\bar b(z)}{\bar a(z)}.
\end{equation}
Generically, it is known that $a(z)$ has a finite number of zeros that lie on the circle $|z|=q_{0}$ (minus the branch points $\pm q_{0}$), and due to the symmetry \eqref{e:ab_sym_1}, if $a(\zeta_{j})=0$ then $\bar{a}(\zeta_{j}^{*})=0$. So, the discrete eigenvalues come in pairs
\begin{equation}
    \label{e:disc_eig}
    \{\zeta_{j},\zeta_{j}^{*}\}_{j=1}^{J},\qquad \zeta_{j}=q_{0}e^{i\alpha_{j}},\qquad 0<\alpha_{j}<\pi.
\end{equation}
At each discrete eigenvalue, \eqref{e:wron} implies that the Jost eigenfunctions become proportional:
\begin{equation}
\phi(x,t,\zeta_{j})=b_{j}\psi(x,t,\zeta_{j}),\qquad \bar\phi(x,t,\zeta_{j}^{*})={b}^{*}_{j}\bar\psi(x,t,\zeta_{j}^{*}),
\end{equation}
for some constant $b_{j}\in \Complex$ satisfying the symmetry $b_{j}^{*}=-b_{j}$. With this in mind, the residue contributions from each pair of discrete eigenvalues are
\begin{equation}
\label{e:res_rh}
    \Res_{z=\zeta_{j}}\frac{\phi(x,t,z)}{a(z)}=C_{j}\psi(x,t,\zeta_{j}),\qquad
    \Res_{z=\zeta_{j}^{*}}\frac{\bar \phi(x,t,z)}{\bar a(z)}=C_{j}^{*}\bar \psi(x,t,\zeta^{*}_{j}),
\end{equation}
where $C_{j}=b_{j}/a^\prime(\zeta_{j})$ is the norming constant associated with the discrete eigenvalue (hereafter, $^\prime$ denotes differentiation with respect to the scattering parameter $z$). Moreover, the first symmetry in \eqref{e:ab_sym_2} gives:
$$
\bar{a}'(z)=-e^{-2i\theta}\frac{q_0^2}{z^2}a'(q_0^2/z),
$$
and therefore the norming constants are such that 
\begin{equation}
\label{e:2ndsymmCj}
C_{j}^{*}=
e^{2i(\theta-\alpha_{j})}
C_{j}.
\end{equation}
Note that, like the scattering coefficients and the reflection coefficients, the norming constants are time-independent because the Jost eigenfunctions have been defined as simultaneous solutions of the Lax pair.

The inverse scattering problem is then solved by formulating \eqref{e:rh} as a RHP across the real $z$ axis. After applying Cauchy projectors to \eqref{e:rh} and accounting for the residues, the RHP for the Jost eigenfunctions $\psi(x,t,z)$ and $\bar{\psi}(x,t,z)$ is converted into the following linear system of algebraic-integral equations:
\bse
\label{e:lin_sys}
\begin{eqnarray}  \psi(x,t,z)e^{-i\Omega(x,t,z)}&=&\begin{bmatrix}
        -iq_{+}/z\\1
\end{bmatrix}+\sum_{j=1}^{J}\frac{e^{-i\Omega(x,t,\zeta_{j}^*)}\bar{\psi}(x,t,\zeta_{j}^{*})C_{j}^{*}}{z-\zeta_{j}^{*}} \\ &&-\,\frac{1}{2\pi  i}\int_{-\infty}^{\infty}\frac{e^{-i\Omega(x,t,\zeta)}\bar \psi(x,t,\zeta)\rho^{*}(\zeta)}{\zeta-(z+i0)}d\zeta, \notag \\
        \bar \psi(x,t,z)e^{i\Omega(x,t,z)}&=&\begin{bmatrix}
        1\\iq_{+}^{*}/z
\end{bmatrix}+\sum_{j=1}^{J}\frac{e^{i\Omega(x,t,\zeta_{j})}\psi(x,t,\zeta_{j})C_{j}}{z-\zeta_{j}}\\
\notag &&+\,\frac{1}{2\pi  i}\int_{-\infty}^{\infty}\frac{e^{i\Omega(x,t,\zeta)}\psi(x,t,\zeta)\rho(\zeta)}{\zeta-(z-i0)}d\zeta.
\end{eqnarray}
\ese
Finally, once the eigenfunctions are known
the potential (i.e., the solution $q(x,t)$ to \eqref{e:NLS}) can be recovered 
from the large-$z$ asymptotic behavior of the eigenfunctions. 

\section{Completeness of squared eigenfunctions for general potentials}
\label{S3}
The main purpose of this section is to prove completeness of the squared eigenfunctions for general potentials.
We first show how to calculate variations of the scatting data in terms of a variation of the potential, and vice versa, which in turn will yield the adjoint squared eigenfunctions and the squared eigenfunctions. Then, we use the variations to prove the completeness result. The methodology follows \cite{Yang}, while also accounting for the nonzero boundary conditions. Throughout this section, we suppress explicit dependence on $t$. 

\subsection{Variation in the scattering data from a variation in the potential}
\label{S3.1}
Since the eigenfunction $\Phi$ defined in \eqref{e:jost_phi} satisfies \eqref{e:lax_x},
a variation  $\delta\Phi$ in the eigenfunction corresponding to a variation  $\delta Q$ in the potential must satisfy
\begin{equation}
\label{e:Phi_var}
    \delta\Phi_{x}=-ik\sigma_{3}\delta\Phi+Q\delta\Phi+(\delta Q)\Phi\,,\qquad \lim_{x\rightarrow-\infty}\delta\Phi=0.
\end{equation}
One can check that the solution of \eqref{e:Phi_var} is given by
\begin{equation}
\label{e:Phi_sol}
    \delta\Phi(x,z)=\Phi(x,z)\int_{-\infty}^{x}\Phi^{-1}(y,z)\delta Q(y)\Phi(y,z)dy.
\end{equation}
Taking the limit as $x\rightarrow+\infty$, we have $\Phi=\Psi S\rightarrow E_{+}e^{-i\Omega\sigma_{3}}S$, so that \eqref{e:Phi_sol} becomes
\begin{equation}
\label{e:S_var}
    \delta S(z)=\int_{-\infty}^{\infty}\Psi^{-1}(y,z)\delta Q(y)\Phi(y,z)dy.
\end{equation}
If we write the eigenfunctions as
\begin{equation}
    \Phi=\begin{bmatrix}
        \phi&\bar\phi
    \end{bmatrix}=\begin{bmatrix}
        \phi_{1}&\bar\phi_{1}\\\phi_{2}&\bar\phi_{2}
    \end{bmatrix},\qquad \Psi=\begin{bmatrix}
        \bar\psi&\psi
    \end{bmatrix}=\begin{bmatrix}
        \bar\psi_{1}&\psi_{1}\\\bar\psi_{2}&\psi_{2}
    \end{bmatrix},
\end{equation}
then their inverses are
\begin{equation}
\label{e:inverses}
    \Phi^{-1}=\frac{1}{\gamma(z)}\begin{bmatrix}
        \hat{\bar\phi}\\\hat\phi
    \end{bmatrix}=\frac{1}{\gamma(z)}\begin{bmatrix}
        \bar\phi_{2}&-\bar\phi_{1}\\
        -\phi_{2}&\phi_{1}
    \end{bmatrix}\,,\qquad \Psi^{-1}=\frac{1}{\gamma(z)}\begin{bmatrix}
        \hat\psi\\\hat{\bar\psi}
    \end{bmatrix}=\frac{1}{\gamma(z)}\begin{bmatrix}
        \psi_{2}&-\psi_{1}\\
        -\bar\psi_{2}&\bar\psi_{1}
    \end{bmatrix},
\end{equation}
where we recall that $\gamma(z)=\det\Phi=\det\Psi=1-q_{0}^{2}/z^{2}$ (cf \eqref{e:gamma_def}).

With these definitions, \eqref{e:S_var} gives entrywise:
\bse
\label{e:ab_var}
\begin{gather}
    \delta a(z)=\frac{1}{\gamma(z)}\int_{-\infty}^{\infty}\hat\psi(y,z)\delta Q(y)\phi(y,z)dy,\quad 
    \delta b(z)=\frac{1}{\gamma(z)}\int_{-\infty}^{\infty}\hat{\bar\psi}(y,z)\delta Q(y)\phi(y,z)dy,\\
    \delta \bar a(z)=\frac{1}{\gamma(z)}\int_{-\infty}^{\infty}\hat{\bar\psi}(y,z)\delta Q(y)\bar\phi(y,z)dy,\quad
    \delta \bar b(z)=\frac{1}{\gamma(z)}\int_{-\infty}^{\infty}\hat\psi(y,z)\delta Q(y)\bar\phi(y,z)dy.
\end{gather}
\ese
The variations in the reflection coefficients defined by $\rho=b/a$ and $\bar\rho=\bar{b}/\bar{a}$ (cf. \eqref{e:ref_def}) are found to be
\begin{subequations}
\label{e:rho_var}
\begin{gather}
    \delta\rho(z)=\frac{1}{\gamma(z)a(z)^{2}}\int_{-\infty}^{\infty}\hat\phi(y,z)\delta Q(y)\phi(y,z)dy, \\
    \delta\bar\rho(z)=\frac{1}{\gamma(z)\bar a(z)^{2}}\int_{-\infty}^{\infty}\hat{\bar \phi}(y,z)\delta Q(y)\bar\phi(y,z)dy,
\end{gather}
\end{subequations}
where \eqref{e:ab_var} as well as $\hat\phi=a\hat{\bar\psi}-b\hat\psi$ and $\hat{\bar\phi}=\bar a \hat\psi-\bar b\hat{\bar\psi}$ have been used. If we define the inner product\footnote{It is customary to define the inner product this way, without explicit complex conjugation, with the implication that the left argument is a member of an appropriate dual space. The adjoint $M^{A}$ of an operator $M$ is defined such that $\langle M^{A}f,g\rangle=\langle f,Mg\rangle$.}
\begin{equation}
\label{e:ip}
    \langle f,g\rangle=\int_{-\infty}^{\infty}f(y)^{T}g(y)dy,
\end{equation}
then \eqref{e:rho_var} can be written as
\begin{gather}
\label{e:var_refl}
    \delta\rho(z)=\frac{1}{\gamma(z)a(z)^{2}}\left\langle\chi,\begin{bmatrix}
            \delta q\\\delta q^{*}
        \end{bmatrix}\right\rangle,\qquad
            \delta\bar\rho(z)=\frac{1}{\gamma(z)\bar a(z)^{2}}\left\langle\bar\chi,\begin{bmatrix}
            \delta q\\\delta q^{*}
        \end{bmatrix}\right\rangle,
\end{gather}
where the so-called \textbf{adjoint squared eigenfunctions} are defined as:
\begin{equation}
\label{e:adj_squared}
    \chi(x,z)=\begin{bmatrix}
        -\phi_{2}(x,z)^{2}\\\phi_{1}(x,z)^{2}
    \end{bmatrix},\qquad \bar\chi(x,z)=\begin{bmatrix}
        \bar\phi_{2}(x,z)^{2}\\-\bar\phi_{1}(x,z)^{2}
    \end{bmatrix}.
\end{equation}
The analyticity properties of the Jost eigenfunctions in Sec~\ref{S2} imply that $\chi(x,z)$ is analytic for $\Im z>0$, and $\bar{\chi}(x,z)$ is analytic for $\Im z<0$. Note also that on account of the symmetries \eqref{e:phi_psi_sym_1}-\eqref{e:ab_sym_1}, we have that the two equations in \eqref{e:var_refl} are equivalent. 
\begin{remark}
Since $\gamma(\pm q_{0})=0$, two possible scenarios can occur: (i) If the unperturbed reflection coefficient is nonzero, then $a(z)$ has simple poles at $z=\pm q_{0}$, so we have $\delta\rho(\pm q_{0})=0$ and the property $|\rho(\pm q_{0})|=1$ is preserved. For the present discussion, we assume this to be the case. (ii) If the unperturbed reflection coefficient is zero, then $a(\pm q_{0})$ is finite, meaning that $\delta\rho$ becomes singular as $z\rightarrow\pm q_{0}$. Notably, this happens in the case of pure solitons, to be discussed later. The implication is that for infinitesimally small variations in the potential $\delta q$, the resulting variation in the reflection coefficient $\delta\rho$ is finite. This causes the generation of a radiation shelf, a phenomenon that is also observed in the perturbation theory associated with the KdV equation, due to a similar singularity that occurs as the spectral parameter approaches zero (see for example \cite{KarpmanZETF1977,HermanJPA1990}). 
\end{remark}
\subsection{Variation in the potential from a variation in the scattering data}
\label{S3.2}
In matrix form, the relations between the Jost functions given in \eqref{e:rh} can be expressed as
\bse
\label{e:rh_mat}
\begin{gather}
    \bar\mu=\mu G,\qquad
    G=\begin{bmatrix}
        1&e^{-2i\Omega(x,z)}\bar\rho(z)\\
        -e^{2i\Omega(x,z)}\rho(z)&1-\rho(z)\bar\rho(z)
    \end{bmatrix}, \qquad z\in \Real,\\
    \mu=\begin{bmatrix}
        e^{i\Omega(x,z)}\displaystyle{\frac{\phi(x,z)}{a(z)}}\  & e^{-i\Omega(x,z)}\psi(x,z)
    \end{bmatrix},\quad \bar\mu=\begin{bmatrix}
       e^{i\Omega(x,z)}\bar{\psi}(x,z)\  & e^{-i\Omega(x,z)}
       \displaystyle{\frac{\bar{\phi}(x,z)}{\bar{a}(z)}}
    \end{bmatrix},
\end{gather}
\ese
which is a RHP across the real $z$-axis. To start with, we assume that $a(z)$ and $\bar{a}(z)$ have no zeros, i.e., $\mu$ and $\bar\mu$ are analytic functions of $z$ in the upper and lower half planes, respectively. A variation of \eqref{e:rh_mat} reads:
\begin{equation}
    \delta\bar\mu=(\delta \mu)\, G+\mu\, \delta G\,, \qquad \delta G=\begin{bmatrix}
        0&e^{-2i\Omega(x,z)} \delta\bar\rho(z) \\-e^{2i\Omega(x,z)}\delta\rho(z)\ \  &-\rho(z)\delta\bar\rho(z)-\bar\rho(z)\delta\rho(z)
    \end{bmatrix}.
\end{equation}
If we write $G=\mu^{-1}\bar\mu$, and introduce the definitions
\begin{equation}
\nu=(\delta\mu)\mu^{-1},\qquad \bar\nu=(\delta\bar\mu) \bar\mu^{-1},\qquad F=-\mu(\delta G) \bar\mu^{-1},
\end{equation}
then we have the RHP: 
\begin{equation}
\label{e:rh_nu}
    \nu-\bar\nu=F.
\end{equation}
After simplification, $F$ is given by
\begin{equation}
\label{e:H}
    F=\Psi\begin{bmatrix}
        0&-\delta\bar\rho(z)\\
        \delta\rho(z)&0
    \end{bmatrix}\Psi^{-1}.
\end{equation}
Using the large-$z$ asymptotic behavior of the Jost eigenfunctions (which can be found, for example, in \cite{P2023}) in their respective half-planes we have that
\begin{equation}
\label{e:nu_asympt}
    \nu,\bar\nu\sim\begin{bmatrix}
       \mathcal O(1/z)&-i\delta q/z\\
       i\delta q^{*}/z&\mathcal O(1/z)
   \end{bmatrix},\qquad|z|\rightarrow\infty.
\end{equation}
Since $\nu,\bar\nu\rightarrow0$ as $|z|\rightarrow\infty$, the solution on \eqref{e:rh_nu} can be written down in terms of Cauchy projectors as
\begin{gather}
\label{e:rh_sol}
    \nu(x,z)=\frac{1}{2\pi i}\int_{-\infty}^{\infty}\frac{F(x,\zeta)}{\zeta-(z+i0)}d\zeta,\qquad
    \bar\nu(x,z)=\frac{1}{2\pi i}\int_{-\infty}^{\infty}\frac{F(x,\zeta)}{\zeta-(z-i0)}d\zeta.
\end{gather}
Comparing the large-$z$ asymptotic behavior of \eqref{e:rh_sol} with \eqref{e:nu_asympt}
it can be deduced that
\begin{equation}
    \label{e:dq_vec}
    \begin{bmatrix}
        \delta q(x)\\
        \delta q^{*}(x)
    \end{bmatrix}=\frac{1}{2\pi}\int_{-\infty}^{\infty}\frac{1}{\gamma(\zeta)}\Big\{\eta(x,\zeta)\delta\rho(\zeta)+\bar\eta(x,\zeta)\delta\bar\rho(\zeta)\Big\}d\zeta,
\end{equation}
where the \textbf{squared eigenfunctions} are defined as:
\begin{equation}
\label{e:squared}
    \eta(x,z)=\begin{bmatrix}
        \psi_{1}(x,z)^{2}\\
        \psi_{2}(x,z)^{2}
    \end{bmatrix},\qquad \bar\eta(x,z)=\begin{bmatrix}
        \bar\psi_{1}(x,z)^{2}\\
        \bar\psi_{2}(x,z)^{2}
    \end{bmatrix}.
\end{equation}
Again, the analyticity properties of the Jost eigenfunctions in Sec~\ref{S2} imply that $\eta(x,z)$ is analytic for $\Im z>0$, and $\bar{\eta}(x,z)$ is analytic for $\Im z<0$.
Moreover, due to the symmetries \eqref{e:phi_psi_sym_2} and \eqref{e:ab_sym_2}, the squared eigenfunctions and reflection coefficients satisfy
\begin{equation}
    \label{e:eta_sym_q0}
    \bar\eta(x,z)=-\frac{q_{0}^{2}}{z^{2}}e^{-2i\theta}\eta(x,q_{0}^{2}/z),\qquad\delta\bar\rho(z)=-e^{2i\theta}\delta\rho(q_{0}^{2}/z),
\end{equation}
which allows one to combine the two terms in \eqref{e:dq_vec} into one by 
making a change of variables $\zeta'=q_{0}^{2}/\zeta$ in the second integral:
\begin{equation}
    \label{e:var_pot}
    \begin{bmatrix}
        \delta q(x)\\
        \delta q^{*}(x)
    \end{bmatrix}=
    \frac{1}{2\pi}\int_{-\infty}^{\infty}\frac{1}{\gamma(\zeta)}\eta(x,\zeta)\delta\rho(\zeta)\left(1-\frac{q_{0}^{2}}{\zeta^{2}}\right)d\zeta=
    \frac{1}{2\pi}\int_{-\infty}^{\infty}\eta(x,\zeta)\delta\rho(\zeta)d\zeta.
\end{equation}
This reduction is in contrast to the case of zero boundary conditions, where both squared eigenfunctions need to be retained since there is no symmetry analogous to \eqref{e:eta_sym_q0}.
\subsection{Completeness relation for general potentials}
\label{S3.3}
As shown in the last section, a variation in the potential induced by a variation in the reflection coefficient can be expressed in terms of the squared eigenfunction $\eta(x,z)$  defined in \eqref{e:squared} via Eq.~\eqref{e:var_pot}. Conversely, a variation in the reflection coefficient induced by a variation in the potential is expressed in terms of the adjoint squared eigenfunction $\chi(x,z)$ in \eqref{e:adj_squared}  via \eqref{e:var_refl}.
Inserting \eqref{e:var_pot} into \eqref{e:var_refl}, we have
\begin{equation}
    \delta\rho(z)=\frac{1}{2\pi\gamma(z)a(z)^{2}}\int_{-\infty}^{\infty}\big\langle\chi(z),\eta(z)\big\rangle\delta\rho(\zeta)d\zeta,
\end{equation}
from which we obtain the inner product between the squared eigenfunction $\eta$ and its adjoint $\chi$:
\begin{equation}
    \label{e:ip_cont}
    \big\langle\chi(z),\eta(\zeta)\big\rangle=2\pi\gamma(z)a(z)^{2}\delta(z-\zeta).
\end{equation}
On the other hand, substituting \eqref{e:var_refl} into \eqref{e:var_pot} and using the definition of the inner product \eqref{e:ip} gives
\begin{equation}
            \begin{bmatrix}
        \delta q(x)\\
        \delta q^{*}(x)
    \end{bmatrix}=\frac{1}{2\pi}\int_{-\infty}^{\infty}\int_{-\infty}^{\infty}\frac{1}{\gamma(\zeta)a(\zeta)^{2}}\eta(x,\zeta)\chi(y,\zeta)^{T}d\zeta\begin{bmatrix}
            \delta q(y)\\\delta q^{*}(y)
        \end{bmatrix}dy,
\end{equation}
which leads to the closure/completeness relation:
\begin{equation}
    \label{e:closure_nopoles}
    \int_{-\infty}^{\infty}\frac{1}{\gamma(\zeta)a(\zeta)^{2}}\eta(x,\zeta)\chi(y,\zeta)^{T}d\zeta=2\pi\delta(x-y)I.
\end{equation}
Recall that both \eqref{e:ip_cont} and \eqref{e:closure_nopoles} have been derived under the assumption that $a(z)$ has no zeros and that the unperturbed reflection coefficient $\rho(z)$ is nonzero. If this is the case, the integrand has no poles, the set $\{\eta(x,z):z\in\mathbb{R}\}$ is complete, and \eqref{e:closure_nopoles} is the closure relation. Generically, $a(z)$ has a finite number of simple zeros $\{z=\zeta_{j},\;\text{Im}\,\zeta_{j}>0\}_{j=1}^{J}$ that lie on the circle $|z|=q_{0}$. Furthermore, in any reflectionless case, poles at $z=\pm q_{0}$ coming from the fact that $\gamma(\pm q_{0})=0$ also must be accounted for. Therefore, a more general closure relation is required. The general closure relation can be obtained by simply replacing the integral in \eqref{e:closure_nopoles} with one taken over a contour in the upper-half $z$-plane from $-\infty+i0$ to $+\infty+i0$ that passes above the circle $|z|=q_{0}$. Furthermore, when discrete eigenvalues are present (i.e., when $a(z)$ has zeros) the orthogonality of the continuous eigenfunctions as given in \eqref{e:ip_cont} still holds for $z,\zeta\in\mathbb{R}$, and will be supplemented with additional orthogonality relations between the discrete squared eigenfunctions that arise due to the residues at the complex zeros of $a(z)$.

The most compact form of the general closure relation is:
\begin{equation}
\label{e:closure_generic}
    \int_{\Gamma}\frac{1}{\gamma(\zeta)a(\zeta)^{2}}\eta(x,\zeta)\chi(y,\zeta)^{T}d\zeta=2\pi\delta(x-y)I,
\end{equation}
where $\Gamma$ is a contour from $-\infty+i0$ to $+\infty+i0$ that passes above the circle of radius $q_{0}$ in the upper half plane. A justification of the generalization of \eqref{e:closure_nopoles} to the contour integral \eqref{e:closure_generic} is provided in Appendix~\ref{SA}. 
Explicitly accounting for the residue contributions from the poles at the discrete eigenvalues $\zeta_{j}$ (given in Appendix~\ref{SA}), the closure relation can be written as 
\begin{eqnarray}
\label{e:closure_poles}
       2\pi\delta(x-y)I &=&\int_{C}\frac{1}{\gamma(\zeta)a(\zeta)^{2}}\eta(x,\zeta)\chi(y,\zeta)^{T}d\zeta\\ \notag  &&-\,2\pi i\sum_{j=1}^{J}\frac{1}{\gamma(\zeta_{j}){a}'(\zeta_{j})^{2}}\Big[\eta(x,\zeta_{j})\Theta(y,\zeta_{j})^{T}+\eta'(x,\zeta_{j})\chi(y,\zeta_{j})^{T}\Big],
\end{eqnarray}
where $C$ is a contour along the real axis indented in the upper-half plane to avoid $\pm q_{0}$, and we have defined
\begin{equation}
\label{e:Theta}
    \Theta(x,\zeta_{j})=\chi'(x,\zeta_{j})+\frac{\gamma(\zeta_{j}){a}''(\zeta_{j})-\gamma'(\zeta_{j}){a}'(\zeta_{j})}{\gamma(\zeta_{j}){a}'(\zeta_{j})}\chi(x,\zeta_{j}).
\end{equation}
As before, prime denotes differentiation with respect to $z$. Note that in the case of zero boundary conditions, the closure relation contains two continuous and four discrete squared eigenfunctions. In the present case, the number of independent eigenfunctions is halved by accounting for the symmetry $z\mapsto q_{0}^{2}/z$.

Finally, we can account for the possibility of simple poles on the real axis at $z=\pm q_{0}$ (which are due to $\gamma(\pm q_0)=0$, and are present whenever $a(\pm q_0)$ is finite, which occurs for reflectionless potentials) by replacing the integral over $C$ with a principal value integral over the real axis, while accounting for the residues
\begin{equation}
\label{e:res_q0}
    \Res_{z=\pm q_0}\frac{\eta(x,z)\chi(y,z)^{T}}{\gamma(z)a(z)^{2}}= \frac{1}{\gamma'(\pm q_{0})a(\pm q_{0})^{2}}\eta(x,\pm q_{0})\chi(y,\pm q_{0})^{T},
\end{equation}
with a factor of $\pi i$. Again, we stress that \eqref{e:res_q0} is only nonzero in the reflectionless case, for which $a(\pm q_{0})$ is finite. In the case of nonzero reflection, where $a(z)$ has poles at $\pm q_{0}$ (see Sec~\ref{S2}), the earlier form of the closure relation \eqref{e:closure_poles} (with integral taken over the real axis rather than $C$) is sufficient. With these residues incorporated, the \textbf{full closure relation} is:
\begin{gather}
       2\pi\delta(x-y)I=\dashint_{-\infty}^{\infty}\frac{1}{\gamma(\zeta)a(\zeta)^{2}}\eta(x,\zeta)\chi(y,\zeta)^{T}d\zeta-\pi i\sum_{\pm}\frac{1}{\gamma'(\pm q_{0})a(\pm q_{0})^{2}}\eta(x,\pm q_{0})\chi(y,\pm q_{0})^{T}\nonumber\\ -\,2\pi i\sum_{j=1}^{J}\frac{1}{\gamma(\zeta_{j}){a}'(\zeta_{j})^{2}}\Big[\eta(x,\zeta_{j})\Theta(y,\zeta_{j})^{T}+\eta'(x,\zeta_{j})\chi(y,\zeta_{j})^{T}\Big].
       \label{e:closure_final}
\end{gather}
Here and throughout the paper, $\dashint$ denotes the Cauchy principal value integral. It is worth noting at this point that while several previous works (including \cite{KonotopPRE1994,ChenJPA1997}) derived essentially equivalent completeness relations, the half residues due to the poles at $\pm q_{0}$ were not explicitly accounted for. The inclusion of these contributions plays an important role in the perturbation theory, on which we will elaborate in Sec~\ref{S6}.

\section{Linearization operator and completeness relation for a single soliton}
\label{S4}
In this section, we discuss the linearization operator of the NLS equation with nonzero boundary conditions and its relation to the squared eigenfunctions, and we derive the explicit expression of the completeness relation for a 1-soliton solution.  
\subsection{The linearization operator}
\label{S4.1}
Upon directly taking a variation of the NLS equation \eqref{e:NLS}, we get
\begin{equation}
\label{e:var_nls}
    i\delta q_{t}(x,t)+\delta q_{xx}(x,t)-2(2|q(x,t)|^{2}-q_{0}^{2})\delta q(x,t)-2q(x,t)^{2}\delta q^{*}(x,t)=0.
\end{equation}
Combining \eqref{e:var_nls} with its complex conjugate, a variation in the potential should satisfy
\begin{equation}
\label{e:var_nls_vec}
    \mathcal{L}\begin{bmatrix}\delta q(x,t)\\\delta q^{*}(x,t)\end{bmatrix}=0,
\end{equation}
where the linearization operator is
\begin{equation}
\label{e:L_x}
    \mathcal{L}=\begin{bmatrix}
        i\partial_{t}+\partial_{xx}-2(2|q(x,t)|^{2}-q_{0}^{2})&-2q(x,t)^{2}\\
        2q^{*}(x,t)^{2}&i\partial_{t}-\partial_{xx}+2(2|q(x,t)|^{2}-q_{0}^{2})
    \end{bmatrix}.
\end{equation}
Inserting \eqref{e:var_pot} into \eqref{e:var_nls_vec} shows that the squared eigenfunction $\eta$ is an eigenfunction of the linearization operator with zero eigenvalue, namely
\begin{equation}
    \label{e:L_eta}
    \mathcal{L}\eta(x,t,z)=0.
\end{equation}
Moreover, the formal adjoint of $\mathcal{L}$ is
\begin{equation}
    \mathcal{L}^{A}=\begin{bmatrix}
        -i\partial_{t}+\partial_{xx}-2(2|q(x,t)|^{2}-q_{0}^{2})&2q^{*}(x,t)^{2}\\
        -2q(x,t)^{2}&-i\partial_{t}-\partial_{xx}+2(2|q(x,t)|^{2}-q_{0}^{2})
    \end{bmatrix},
\end{equation}
and from
\begin{equation}
    \langle\chi(z),\mathcal{L}\eta(\zeta)\rangle=\langle\mathcal{L}^{A}\chi(z),\eta(\zeta)\rangle=0,
\end{equation}
we deduce that the adjoint squared eigenfunction $\chi$ satisfies
\begin{equation}
\label{e:LA_chi}
    \mathcal{L}^{A}\chi(x,t,z)=0.
\end{equation}
Apart from an arbitrary phase, the exact single dark soliton solution of \eqref{e:NLS} is
\begin{equation}
\label{e:1_sol}
    q(x,t)=u(\xi)=k_{1}+i\Lambda_{1}\tanh\xi,\qquad \xi\equiv\Lambda_{1}(x+2k_{1}t-x_{1}),
\end{equation}
where $\zeta_{1}=k_{1}+i\Lambda_{1}=q_{0}e^{i\theta}$. Thus, when performing a linearization around the 1-soliton solution, it proves convenient to express the linearization operator and its adjoint in terms of the traveling coordinate $\xi$ rather that $x$ (while allowing for potential explicit dependence on $t$). In this case, we have $\partial_{t}\rightarrow\partial_{t}+2\Lambda_{1}k_{1}\partial_{\xi}$, $\partial_{x}\rightarrow\Lambda_{1}\partial_{\xi}$ and 
\begin{equation}
\label{e:L_split}
    \mathcal{L}=i\partial_{t}+L,\qquad \mathcal{L}^{A}=-i\partial_{t}+L^{A},
\end{equation}
where
\bse
\label{e:L_xi}
\begin{gather}
L=\begin{bmatrix}
\Lambda_{1}^{2}\partial_{\xi}^{2}+2i\Lambda_{1}k_{1}\partial_{\xi}-2(2|u(\xi)|^{2}-q_{0}^2)&-2u(\xi)^{2}\\
        2u^{*}(\xi)^{2}&-\Lambda_{1}^{2}\partial_{\xi}^{2}+2i\Lambda_{1}k_{1}\partial_{\xi}+2(2|u(\xi)|^{2}-q_{0}^{2})
    \end{bmatrix},\\
    L^{A}=\begin{bmatrix}
\Lambda_{1}^{2}\partial_{\xi}^{2}-2i\Lambda_{1}k_{1}\partial_{\xi}-2(2|u(\xi)|^{2}-q_{0}^2)&2u(\xi)^{*2}\\
        -2u(\xi)^{2}&-\Lambda_{1}^{2}\partial_{\xi}^{2}-2i\Lambda_{1}k_{1}\partial_{\xi}+2(2|u(\xi)|^{2}-q_{0}^{2})
    \end{bmatrix}.
\end{gather}
\ese
In the single soliton case, the squared eigenfunctions can be computed explicitly by solving the linear system \eqref{e:lin_sys} 
for the Jost eigenfunctions $\psi(x,t,z)$ and $\bar{\psi}(x,t,z)$ with $J=1$, $\rho(z)\equiv0$, and discrete eigenvalue and norming constant given by:
\begin{equation}
\zeta_{1}=k_{1}+i\Lambda_{1}, \qquad     C_{1}=-2\Lambda_{1}e^{2\Lambda_{1}x_{1}},
\end{equation}
and $\phi(x,t,z)$ can be recovered from the above via:
\begin{equation}
\label{e:phi_psi}
 \phi(x,t,z)=a(z)\bar{\psi}(x,t,z), 
\end{equation}
(cf. \eqref{e:rh} with $\rho(z)\equiv 0$) as well as the explicit expression for $a(z)$ in the 1-soliton case, namely:
\begin{equation}
\label{e:a1sol}
a(z)=\frac{z-\zeta_1}{z-\zeta_1^*}.
\end{equation}
Eqs.~\eqref{e:squared} and \eqref{e:adj_squared} then give the following explicit expression for the squared eigenfunctions: 
\bse
\begin{eqnarray}
\eta(x,t,z)&=&\begin{bmatrix}
        [iq_{+}z^{-1}(z-\zeta_{1}^{*})+\Lambda_{1}e^{-\xi}\sech\xi]^2\\
    [(z-\zeta_{1}^{*})-i\Lambda_{1}e^{-\xi}\sech\xi]^2\end{bmatrix}\frac{e^{2i\Omega(x,t,z)}}{(z-\zeta_{1}^{*})^{2}},\\
    \chi(x,t,z)&=&\begin{bmatrix}
    -[iq_{-}z^{-1}(z-\zeta_{1})-\Lambda_{1}e^{-\xi}\sech\xi]^2\\
     [(z-\zeta_{1})+i\Lambda_{1} e^{-\xi}\sech\xi]^2
\end{bmatrix}\frac{e^{-2i\Omega(x,t,z)}}{(z-\zeta_{1}^{*})^{2}}.
\end{eqnarray}
\ese
A detailed derivation of the 1-soliton Jost eigenfunctions from the linear system is provided in Appendix~\ref{SB}. In the traveling reference frame, we can define the corresponding time-independent squared eigenfunctions
\bse
\label{e:ZY_def}
\begin{eqnarray}
\label{e:Z_def}
    \mathcal{Z}(\xi,z)&=&\eta(x,t,z)e^{-4i\lambda(k-k_{1})t-2i\lambda x_{1}}(z-\zeta_{1}^{*})^{2},\\
\label{e:Y_def}
\Upsilon(\xi,z)&=&\chi(x,t,z)e^{4i\lambda(k-k_{1})t+2i\lambda x_{1}}(z-\zeta_{1}^{*})^{2},
\end{eqnarray}
\ese
where the $z$-dependent denominators have been removed to simplify later calculations, and $k=k(z)$ and $\lambda=\lambda(z)$ are as defined in \eqref{e:uni}.  Straightforward calculations yield the explicit forms:
\bse
\label{e:ZY_exp}
\begin{eqnarray}
\label{e:Z_exp}
\mathcal{Z}(\xi,z)&=&\begin{bmatrix}
        [iq_{+}z^{-1}(z-\zeta_{1}^{*})+\Lambda_{1}e^{-\xi}\sech\xi]^2\\
    [(z-\zeta_{1}^{*})-i\Lambda_{1}e^{-\xi}\sech\xi]^2\end{bmatrix}e^{2i\frac{\lambda(z)}{\Lambda_{1}}\xi},\\
\label{e:Y_exp}
    \Upsilon(\xi,z)&=&\begin{bmatrix}
    -[iq_{-}z^{-1}(z-\zeta_{1})-\Lambda_{1}e^{-\xi}\sech\xi]^2\\
     [(z-\zeta_{1})+i\Lambda_{1} e^{-\xi}\sech\xi]^2
\end{bmatrix}e^{-2i\frac{\lambda(z)}{\Lambda_{1}}\xi}.
\end{eqnarray}
\ese
From \eqref{e:ip_cont}, the inner product between the new eigenfunctions is given by:
\begin{equation}
    \label{e:ip_cont_ZY}
    \langle\Upsilon(z),\mathcal{Z}(\zeta)\rangle=\frac{1}{\Lambda_{1}}\int_{-\infty}^{\infty}\Upsilon(\xi',z)^{T}\mathcal{Z}(\xi',\zeta)d\xi'=2\pi\gamma(z)|z-\zeta_{1}|^{4}\delta(z-\zeta).
\end{equation}
Note that the division by $\Lambda_{1}$ is necessary because, here and throughout the paper, the inner product is taken with respect to $x$ and not $\xi$. Substituting \eqref{e:Z_def} into \eqref{e:L_eta} and \eqref{e:Y_def} into \eqref{e:LA_chi} with \eqref{e:L_split} in mind, we find that $\mathcal{Z}$ and $\Upsilon$ are eigenfunctions of $L$ and $L^{A}$, respectively:
\begin{gather}
\label{e:ZY_eig}
    L\mathcal{Z}(\xi,z)=4\lambda(k-k_{1})\mathcal{Z}(\xi,z),\qquad
    L^{A}\Upsilon(\xi,z)=4\lambda(k-k_{1})\Upsilon(\xi,z).
\end{gather}
The discrete eigenfunctions associated with the soliton eigenvalue $\zeta_{1}$ can be obtained by directly evaluating \eqref{e:ZY_exp} at $z=\zeta_1$, from which we get
\begin{gather}
\label{e:ZY_dis}
\mathcal{Z}_1(\xi):=   \mathcal{Z}(\xi,\zeta_{1})=\Lambda_{1}^{2}\begin{bmatrix}
        1\\-1
    \end{bmatrix}\sech^{2}\xi,\qquad
    \Upsilon_1(\xi):=\Upsilon(\xi,\zeta_{1})=-\Lambda_{1}^{2}\begin{bmatrix}
        1\\1
    \end{bmatrix}\sech^{2}\xi.
\end{gather}
Furthermore, setting $z=\zeta_{1}$ in \eqref{e:ZY_eig} shows that 
\begin{equation}
        L\mathcal{Z}_1(\xi)=
    L^{A}\Upsilon_1(\xi)=0.
\end{equation}
Differentiating \eqref{e:ZY_exp} with respect to $z$, the generalized discrete eigenfunctions are found to be
\bse
\begin{gather}
    \mathcal{Z}^\prime_1(\xi):=\left. \partial_{z}\mathcal{Z}(\xi,z)\right|_{z=\zeta_1}
    =-2i\Lambda_{1}\left\{\begin{bmatrix}q_{0}^{2}\zeta_{1}^{-2}\\-1\end{bmatrix}e^{-\xi}\sech\xi-k_{1}\zeta_{1}^{-1}\begin{bmatrix}
        1\\-1
    \end{bmatrix}\xi\sech^{2}\xi\right\},\\
     \Upsilon^\prime_1(\xi) :=\left.  \partial_{z}\Upsilon(\xi,z)\right|_{z=\zeta_1}
     =2i\Lambda_{1}\left\{\begin{bmatrix}q_{0}^{2}\zeta_{1}^{-2}\\1\end{bmatrix}e^{\xi}\sech\xi+k_{1}\zeta_{1}^{-1}\begin{bmatrix}
            1\\1
        \end{bmatrix}\xi\sech^{2}\xi\right\},
\end{gather}
\ese
and differentiating \eqref{e:ZY_eig} shows that they satisfy
\begin{gather}
\label{e:ZY_disc_eig}
    L\mathcal{Z}^\prime_1(\xi)=-4\Lambda_{1}^{2}\zeta_{1}^{-1}\mathcal{Z}_1(\xi),\qquad 
L^{A}\Upsilon^\prime_1(\xi)=-4\Lambda_{1}^{2}\zeta_{1}^{-1}\Upsilon_1(\xi).
\end{gather}
It is worth noting that the identification of the discrete modes $\mathcal{Z}_1$ and ${\mathcal{Z}}_1'$ explicitly resolves the algebraic multiplicity of the zero eigenvalue of the linearized operator $L$. The eigenfunction $\mathcal{Z}_1$ corresponds to the Goldstone mode from translational invariance, while the generalized eigenfunction ${\mathcal{Z}}_1'$ corresponds to the mode from scaling invariance (see for example \cite{Bilas}), confirming that the zero eigenvalue is indeed of algebraic multiplicity two.

\subsection{1-soliton completeness relation}
\label{S4.2}
Here we specialize the completeness relation to the 1-soliton case. Returning to the generic completeness relation \eqref{e:closure_generic} and substituting \eqref{e:ZY_def} as well as \eqref{e:a1sol} for $a(z)$, we have
\begin{equation}
\label{e:closure_generic_ZY}
    \int_{\Gamma}\frac{1}{\gamma(\zeta)|\zeta-\zeta_{1}|^{4}}\mathcal{Z}(\xi,\zeta)\Upsilon(\xi',\zeta)^{T}d\zeta \\ =2\pi\delta(\xi-\xi')I.
\end{equation}
After calculating the residue due to $\zeta_{1}$, the completeness relation can be written as
\begin{eqnarray}
       2\pi\delta(\xi-\xi')I&=&\int_{C}\frac{1}{\gamma(\zeta)|\zeta-\zeta_{1}|^{4}}\mathcal{Z}(\xi,\zeta)\Upsilon(\xi',\zeta)^{T}d\zeta\notag\\&&+\,\frac{\pi\zeta_{1}}{4\Lambda_{1}^{3}} \Big[\mathcal{Z}_1(\xi)\Xi_1(\xi')^{T}+\mathcal{Z}^\prime_1(\xi)\Upsilon_1(\xi')^{T}\Big],
       \label{e:closure_poles_ZY}
\end{eqnarray}
where, as before, $C$ is indented in the upper half plane to avoid $\pm q_{0}$, and we have defined the new generalized eigenfunction
\begin{equation}
\label{e:Xi}
    \Xi_1(\xi)=\Upsilon_1^\prime(\xi)+\frac{2ik_{1}}{\Lambda_{1}\zeta_{1}}\Upsilon_1(\xi).
\end{equation}
One can verify directly that
$\langle\Xi_1(\xi),\mathcal{Z}^\prime_1(\xi)\rangle=0$.
Now, we account for the residues at the branch points. Setting $z=\pm q_{0}$ in \eqref{e:ZY_eig} and noting that $\lambda(\pm q_{0})=0$ shows that
\begin{equation}
    L\mathcal{Z}^\pm_0(\xi)=L^{A}\Upsilon^\pm_0(\xi)=0, \qquad 
    \mathcal{Z}^\pm_0(\xi):=\mathcal{Z}(\xi,\pm q_{0}), \quad
    \Upsilon_0^\pm(\xi):=\Upsilon(\xi, \pm q_0).
\end{equation}
A straightforward calculation shows that the squared eigenfunctions evaluated at $\pm q_{0}$ can be written in terms of the dark soliton $u(\xi)$ as defined in \eqref{e:1_sol} as well as the discrete eigenfunctions given in \eqref{e:ZY_dis} in the following way:
\begin{gather}
\label{e:ZY_q0}
    \mathcal{Z}_0^\pm(\xi)=2(k_{1}\mp q_{0})\begin{bmatrix}
    -u(\xi)\\u^{*}(\xi)\end{bmatrix}-\mathcal{Z}_1(\xi),\qquad 
        \Upsilon_0^\pm (\xi)=2(k_{1}\mp q_{0})\begin{bmatrix}
    u^{*}(\xi)\\u(\xi)\end{bmatrix}-\Upsilon_1(\xi).
\end{gather}
Pulling out the residue contribution to the integral in \eqref{e:closure_poles_ZY} due to $\pm q_{0}$, the 1-soliton completeness relation can be written as:
\begin{eqnarray}
       2\pi\delta(\xi-\xi')I&=&\dashint_{-\infty}^{\infty}\frac{1}{\gamma(\zeta)|\zeta-\zeta_{1}|^{4}}\mathcal{Z}(\xi,\zeta)\Upsilon(\xi',\zeta)^{T}d\zeta \nonumber\\
       &&-\,{\pi i}\sum_{\pm}\frac{\pm1}{{8q_{0}}(k_{1}\mp q_{0})^{2}}\mathcal{Z}_0^\pm(\xi)\Upsilon_0^\pm(\xi')^{T} \nonumber \\
       &&+\,\frac{\pi\zeta_{1}}{4\Lambda_{1}^{3}} \Big[\mathcal{Z}_1(\xi)\Xi_1(\xi')^{T}+\mathcal{Z}^\prime_1(\xi)
       \Upsilon_1(\xi')^{T}\Big]. \label{e:closure_full_ZY}
\end{eqnarray}
Although it may appear that the contributions from $\pm q_{0}$ generate additional discrete eigenfunctions, it is actually the case that these contributions modify the discrete eigenfunctions already present in the completeness relation. Indeed, using \eqref{e:ZY_q0}, we find that
\begin{equation}
\label{e:res_q0_exp}
    \sum_{\pm}\frac{\pm1}{{8q_{0}}(k_{1}\mp q_{0})^{2}}\mathcal{Z}_0^\pm(\xi)\Upsilon_0^\pm(\xi')^{T}=\frac{1}{2\Lambda_{1}^{2}}\Bigg\{\begin{bmatrix}
        -u(\xi)\\u^{*}(\xi)
    \end{bmatrix}\Upsilon_1(\xi')^{T} 
     +\mathcal{Z}_1(\xi)\begin{bmatrix}
        u^{*}(\xi')\\u(\xi')
    \end{bmatrix}^{T}+\frac{k_{1}}{\Lambda_{1}^{2}}\mathcal{Z}_1(\xi)\Upsilon_1(\xi')^{T}\Bigg\}.
\end{equation}
 The last term in \eqref{e:res_q0_exp} can be combined with the $\mathcal{Z}_1\Xi_1^{T}$ term in \eqref{e:closure_full_ZY}. Specifically, we have
\begin{equation}
    \frac{\pi\zeta_{1}}{4\Lambda_{1}^{3}}\mathcal{Z}_1\Xi_1^{T}-\frac{\pi k_{1}}{2\Lambda_{1}^{4}}\mathcal{Z}_1\Upsilon_1^{T}=\frac{\pi\zeta_{1}}{4\Lambda_{1}^{3}}\mathcal{Z}_1\Upsilon_1^{\prime\, T},
\end{equation}
which then gives the completeness relation in the form:
\begin{eqnarray}
       2\pi\delta(\xi-\xi')I&=&\dashint_{-\infty}^{\infty}\frac{1}{\gamma(\zeta)|\zeta-\zeta_{1}|^{4}}\mathcal{Z}(\xi,\zeta)\Upsilon(\xi',\zeta)^{T}d\zeta \notag \\
    &&-\,\frac{\pi i}{2\Lambda_{1}^{2}}\left\{\begin{bmatrix}
        -u(\xi)\\u^{*}(\xi)
    \end{bmatrix}\Upsilon_1(\xi')^{T} +\mathcal{Z}_1(\xi)\begin{bmatrix}
        u^{*}(\xi')\\u(\xi')
    \end{bmatrix}^{T}\right\}\notag\\&&+\,\frac{\pi\zeta_{1}}{4\Lambda_{1}^{3}} \Big[\mathcal{Z}_1(\xi)\Upsilon_1^\prime(\xi')^{ T}+\mathcal{Z}_1^\prime(\xi)\Upsilon_1(\xi')^{T}\Big].
       \label{e:closure_full_ZY_2}
\end{eqnarray}
Finally, the remaining two terms in braces can be absorbed into the $\mathcal{Z}_1\Upsilon_1^{\prime\, T}$ and $\mathcal{Z}^\prime_1\Upsilon^{T}_1$ terms. In particular, if we define
\begin{gather}
    \tilde{\mathcal{Z}}_1(\xi)=\mathcal{Z}_1^\prime(\xi)-\frac{2i\Lambda}{\zeta_{1}}\begin{bmatrix}
        -u(\xi)\\u^{*}(\xi)
    \end{bmatrix},\qquad 
    \label{e:up_tilde}
    \tilde{\Upsilon}_1(\xi)=\Upsilon_1^\prime(\xi)-\frac{2i\Lambda_{1}}{\zeta_{1}}\begin{bmatrix}
        u^{*}(\xi)\\u(\xi)
    \end{bmatrix},
\end{gather}
then the final expression for the \textbf{1-soliton closure relation} is:
\begin{eqnarray}
     2\pi\delta(\xi-\xi')I&=&\dashint_{-\infty}^{\infty}\frac{1}{\gamma(\zeta)|\zeta-\zeta_{1}|^{4}}\mathcal{Z}(\xi,\zeta)\Upsilon(\xi',\zeta)^{T}d\zeta\notag\\&&+\,\frac{\pi\zeta_{1}}{4\Lambda_{1}^{3}} \Big[\mathcal{Z}_1(\xi)\tilde\Upsilon_1(\xi')^{T}+\tilde{\mathcal{Z}}_1(\xi)\Upsilon_1(\xi')^{T}\Big].
       \label{e:closure_full_ZY_3}
\end{eqnarray}
The new generalized eigenfunction $\tilde{\mathcal{Z}}_{1}$ is a linear combination of the Goldstone modes from scaling invariance and from phase invariance. Note that the discrete eigenfunctions satisfy the following orthogonality conditions (which can either be read off from the closure relation or verified directly):
\begin{equation}   \langle\Upsilon_{1},\mathcal{Z}_{1}\rangle=\langle\tilde\Upsilon_{1},\tilde{\mathcal{Z}}_{1}\rangle=0,\qquad\langle\tilde\Upsilon_{1},\mathcal{Z}_{1}\rangle=\langle\Upsilon_{1},\tilde{\mathcal{Z}}_{1}\rangle=8\Lambda_{1}^{3}\zeta_{1}^{-1}.
\end{equation}
\section{Perturbation theory}
\label{S5}
Consider the defocusing NLS equation with an additional small forcing pertubation:
\begin{equation}
\label{e:NLS_pert}
    iq_{t}(x,t)+q_{xx}(x,t)-2(|q(x,t)|^{2}-q_{0}^{2})q(x,t)=\varepsilon F[q(x,t)], \qquad 0< \varepsilon \ll 1,
\end{equation}
and introduce the slow time scale $T=\varepsilon t$ so that $\partial_{t}\rightarrow\partial_{t}+\varepsilon\partial_{T}$.
\begin{remark}
    An exact dark soliton solution of \eqref{e:NLS_pert} with $\varepsilon=0$ has the asymptotic behavior $q(x,t)\rightarrow q_{0}e^{i\theta_{\pm}}$ as $x\rightarrow\pm\infty$ with generic phases $\theta_{\pm}$. Following the technique of \cite{FrantzPRSA2011,CFHKP2025}, substituting this into \eqref{e:NLS_pert} in the limit $x\rightarrow\pm\infty$ while allowing $q_{0}$ and $\theta_{\pm}$ to depend on $T$ yields
    \begin{equation}
        q_{0T}=\Im\left\{e^{-i\theta_{\pm}}F[q_{0}e^{i\theta_{\pm}}]\right\},\qquad\theta_{\pm T}=\mp\frac{1}{q_{0}}\Re\left\{e^{-i\theta_{\pm}}F[q_{0}e^{i\theta_{\pm}}]\right\}.
    \end{equation}
    If the perturbation is such that $F[q_{0}e^{i\theta_{\pm}}]=F[q_{0}]e^{i\theta_{\pm}}$, then we have
    \begin{equation}
        \label{e:bg}
        q_{0T}=\Im F[q_{0}],\qquad (\Delta\theta)_{T}=0,
    \end{equation}
    where $\Delta\theta=\theta_{+}-\theta_{-}$ is the asymptotic phase difference. The first formula in \eqref{e:bg} can be used to determine the evolution of the background amplitude of a perturbed dark soliton \textit{a priori}. In the method outlined below, one instead determines the parameters $k_{1}$ and $\Lambda_{1}$ as in \eqref{e:1_sol} and recovers the background through $q_{0}=\sqrt{k_{1}^{2}+\Lambda_{1}^{2}}$, which necessarily will give the same result as \eqref{e:bg}. 
\end{remark}

When $\varepsilon=0$, the exact dark soliton solution satisfying the boundary conditions in \eqref{e:BC_pm} is $q(x,t)=u(\xi)$ given by \eqref{e:1_sol}. Now, suppose that the soliton parameters depend on the slow time scale:
\begin{equation}
    k_{1}=k_{1}(T),\qquad \Lambda_{1}=\Lambda_{1}(T),\qquad x_{1}=x_{1}(T).
\end{equation}
In this case, we redefine the traveling coordinate $\xi$ as
\begin{equation}
    \label{e:new_xi}
  \xi=\Lambda_{1}(T)\left(x+2\int_{0}^{t}k_{1}\,ds-x_{1}(T)\right).
\end{equation}
Note that since the slow-time-dependent velocity must be integrated, it is actually the case that the dependence of $k_{1}$ on an even slower time scale $\varepsilon^2 t$ could contribute nontrivially to the soliton center. Thus, to ensure 
a correct prediction for the soliton center, one would need to include another time scale in the perturbation theory. This is discussed in the context of a specific example in Remark 3 of Sec~\ref{S7.3}, but is beyond the scope of this work.

Furthermore, to account for slow evolution of the overall phase, we introduce another parameter
\begin{equation}
    \sigma_{1}=\sigma_{1}(T),\qquad \sigma_{1}(0)=0,
\end{equation}
and let
\begin{equation}
\label{e:approx}
    q(x,t)\approx e^{i\sigma_{1}}\big[u(\xi)+\varepsilon\tilde{q}(\xi,t)\big],
\end{equation}
where $\tilde{q}$ is the first order correction and explicit dependence on $T$ is omitted for brevity. We will later show that the slowly evolving parameter $\sigma_{1}(T)$ corresponds to inevitable secular growth in the first order correction, which can be absorbed into the phase at least for small times.
Substituting \eqref{e:approx} into \eqref{e:NLS_pert} leads at $\mathcal{O}(\varepsilon)$ to the following equation:
\begin{subequations}
\begin{gather}
\label{e:AW}
    (i\partial_{t}+L)A(\xi,t)=W(\xi),\qquad A=\begin{bmatrix}
       \tilde q\\\tilde{q}^{\,*}
    \end{bmatrix},\qquad W=\begin{bmatrix}
        w[u]\\
        -w[u]^*
    \end{bmatrix},\\ 
       w[u]=e^{-i\sigma_{1}}F[ue^{i\sigma_{1}}]-ie^{-i\sigma_{1}}\partial_{T}(ue^{i\sigma_{1}}),
\end{gather}
\end{subequations}
where the linearization operator $L$ is the same as in \eqref{e:L_xi}. Note that 
\begin{gather}
\label{e:u_T}
    e^{-i\sigma_{1}}\partial_{T}(ue^{i\sigma_{1}})=k_{1T}-\sigma_{1T}\Lambda_{1}\tanh\xi+i\Lambda_{1T}(\tanh\xi+\xi \sech^{2}\xi) \\ -ix_{1T}\Lambda_{1}^{2}\sech^2\xi+i\sigma_{1T}k_{1}. \notag
\end{gather}
Throughout, the subscript $T$ denotes the partial derivative with respect to $T=\varepsilon t$. According to the fully reduced 1-soliton closure relation \eqref{e:closure_full_ZY_3}, we can expand $A$ and $W$ in terms of the complete set of squared eigenfunctions $\{\mathcal{Z}(\xi,\zeta),\mathcal{Z}_1(\xi),\tilde{\mathcal{Z}}_1(\xi)\}$ with respective adjoint eigenfunctions $\{\Upsilon(\xi,\zeta),\Upsilon_1(\xi),\tilde\Upsilon_1(\xi)\}$ as follows:
\bse
\label{e:AW_exp}
\begin{equation}
        \label{e:W_exp}
        W(\xi)=\frac{\zeta_{1}}{8\Lambda_{1}^{3}}\left[\langle\tilde\Upsilon_{1},W\rangle\mathcal{Z}_1(\xi)+\langle\Upsilon_1,W\rangle\tilde{\mathcal{Z}}_1(\xi)\right]+\frac{1}{2\pi}\dashint_{-\infty}^{\infty}\frac{\langle\Upsilon(\zeta),W\rangle}{\gamma(\zeta)|\zeta-\zeta_{1}|^{4}}\mathcal{Z}(\xi,\zeta)d\zeta,
        \end{equation}\begin{equation}
        \label{e:A_exp}
        A(\xi,t)=\frac{\zeta_{1}}{8\Lambda_{1}^{3}}\left[\langle\tilde\Upsilon_{1},A(t)\rangle\mathcal{Z}_1(\xi)+\langle\Upsilon_1,A(t)\rangle\tilde{\mathcal{Z}}_1(\xi)\right]+\frac{1}{2\pi}\dashint_{-\infty}^{\infty}\frac{\langle\Upsilon(\zeta),A(t)\rangle}{\gamma(\zeta)|\zeta-\zeta_{1}|^{4}}\mathcal{Z}(\xi,\zeta)d\zeta.
\end{equation}
\ese
Substituting \eqref{e:AW_exp} into \eqref{e:AW} and equating coefficients of the discrete eigenfunctions gives
\begin{gather}
\label{e:disc_ev}
i\partial_{t}\langle\Upsilon_1,A(t)\rangle=\langle\Upsilon_1,W\rangle,\qquad
    i\partial_{t}\langle\tilde\Upsilon_{1},A(t)\rangle-4\Lambda_{1}^{2}\zeta_{1}^{-1}\langle\Upsilon_1,A(t)\rangle =\langle\tilde\Upsilon_1,W\rangle.
\end{gather}
To suppress secular growth in \eqref{e:disc_ev}, we need to enforce the orthogonality conditions 
\begin{equation}
    \label{e:orth_cond_1}
    \langle\Upsilon_1,W\rangle=\langle\tilde\Upsilon_1,W\rangle=0.
\end{equation}
The first orthogonality condition above 
yields the evolution of the soliton velocity explicitly in terms of the perturbation:
\begin{equation}
\label{e:k_ev}
    k_{1T}=\frac{1}{2}\,\text{Im}\int_{-\infty}^{\infty}F_{0}\sech^{2}\xi\,d\xi, \qquad F_{0}=e^{-i\sigma_{1}}F[ue^{i\sigma_{1}}].
\end{equation}
As we will show with several examples in Sec~\ref{S7}, once the perturbation is specified, the second orthogonality condition in \eqref{e:orth_cond_1} yields two separate conditions on the soliton parameters (one corresponding to eliminating the divergent integrals). One of these conditions can be used to fully determine the dependence of $\Lambda_{1}$ on the slow time $T$, and the other relates the evolution of the soliton phase and center, $\sigma_{1}$ and $x_{1}$.

Moreover, equating coefficients of the continuous eigenfunctions gives
\begin{equation}
    \label{e:cont_ev}
    i\partial_{t}\langle\Upsilon(\zeta),A(t)\rangle+4\lambda(k-k_{1})\langle\Upsilon(\zeta),A(t)\rangle=\langle\Upsilon(\zeta),W\rangle.
\end{equation}
The solution of \eqref{e:cont_ev} with zero initial condition is
\begin{equation}
    \langle\Upsilon(\zeta),A(t)\rangle=\frac{\langle\Upsilon(\zeta),W\rangle}{4\lambda(k-k_{1})}\left[1-e^{4i\lambda(k-k_{1})t}\right].
\end{equation}
Using the above equation in \eqref{e:A_exp} yields for the first order correction:
\begin{equation}
    \label{e:correction}
    A(\xi,t)=\frac{1}{2\pi}\dashint_{-\infty}^{\infty}\frac{1}{\gamma(\zeta)|\zeta-\zeta_{1}|^{4}}\frac{1-e^{4i\lambda(\zeta)(k(\zeta)-k_{1})t}}{4\lambda(\zeta)\big(k(\zeta)-k_{1}\big)}\langle\Upsilon(\zeta),W\rangle\mathcal{Z}(\xi,\zeta)d\zeta.
\end{equation}
Note that $\gamma(\zeta)=2\zeta^{-1}\lambda(\zeta)$, so this could be written as
\begin{equation}
    \label{e:correction_2}
    A(\xi,t)=\frac{1}{2\pi}\dashint_{-\infty}^{\infty}\frac{\zeta}{|\zeta-\zeta_{1}|^{4}}\frac{1-e^{4i\lambda(\zeta)(k(\zeta)-k_{1})t}}{8\lambda(\zeta)^2\big(k(\zeta)-k_{1}\big)}\langle\Upsilon(\zeta),W\rangle\mathcal{Z}(\xi,\zeta)d\zeta,
\end{equation}
which, up to an inessential difference in the normalization of the squared eigenfunctions, has the same form as the one given in \cite{ChenJPA1997}, with the important caveat that the integral has to be considered in the principal value sense, due to the singularities at $\zeta=\pm q_0$ which arise from $\lambda(\zeta)$.  We stress that due to the fact that the integrand has poles at $\pm q_0$, it is crucial for this integral to be considered in the principal value sense in order for it to be well-defined. This also explains why one has to explicitly single out the contributions at $\pm q_0$ from the closure relation.

It is worth highlighting that the issue of the poles at $\pm q_{0}$ was explicitly acknowledged in \cite{KonotopPRE1994}. Our present approach to account for these poles is to modify the adjoint discrete squared eigenfunctions appropriately (i.e., enforce orthogonality with $\tilde\Upsilon_{1}$ rather than with $\Upsilon'_{1}$) such that the first order correction is reduced to a principal value integral. On the other hand, the approach of \cite{KonotopPRE1994} was to enforce the additional orthogonality condition $\langle\Upsilon_{0}^{\pm},W\rangle=0$. However, this condition is too strong. In fact, we will later show that the inner product $\langle\Upsilon_{0}^{\pm},W\rangle$ is generically nonzero, a fact that is crucial to understanding the generation of the shelf.

\section{First order correction, shelf,  and evolution of the phase}
\label{S6}
As we will show in this section, the first order correction can be effectively used to describe the shelf that develops on each side of the soliton, which in fact results from the singularities
of the scattering data at the points $\pm q_0$, and to obtain estimates for the height and velocity of the shelf. Furthermore, the asymptotic behavior of the first order correction provides the spatial phase gradient on each side of the shelf, as well as a formula for the slow time evolution
of the core soliton phase, which in turn allows one to determine the dependence of the soliton center on the slow time $T$. 

Taking the first component of \eqref{e:correction_2}, on account of \eqref{e:approx} and \eqref{e:AW} the first order correction $\tilde{q}$ is given by:
\begin{equation}
    \label{e:correction_3}
    \tilde{q}(\xi,t)=\frac{1}{2\pi}\dashint_{-\infty}^{\infty}\frac{\zeta}{|\zeta-\zeta_{1}|^{4}}\frac{1-e^{4i\lambda(\zeta)(k(\zeta)-k_{1})t}}{8\lambda(\zeta)^2\big(k(\zeta)-k_{1}\big)}\langle\Upsilon(\zeta),W\rangle \Big[i(\zeta_{1}-q_{0}^{2}\zeta^{-1})+\Lambda_{1}e^{-\xi}\sech\xi\Big]^2 e^{2i\frac{\lambda(\zeta)}{\Lambda_{1}}\xi}d\zeta. 
\end{equation}
First, we study the first order correction on the right side of the shelf, i.e., the asymptotic limit $\xi\rightarrow+\infty$:
\begin{equation}
    \label{e:correction_lim_2}
    \tilde{q}^{+}\sim-\frac{1}{2\pi}\dashint_{-\infty}^{\infty}\frac{\zeta}{|\zeta-\zeta_{1}|^{4}}\frac{1-e^{4i\lambda(\zeta)(k(\zeta)-k_{1})t}}{8\lambda(\zeta)^2\big(k(\zeta)-k_{1}\big)}\langle\Upsilon(\zeta),W\rangle\big(\zeta_{1}-q_{0}^{2}\zeta^{-1}\big)^2 e^{2i\frac{\lambda(\zeta)}{\Lambda_{1}}\xi}d\zeta.
\end{equation}
For large $\xi$, the $\xi$-dependent exponential in \eqref{e:correction_lim_2} is rapidly oscillating, so the primary contributions to the principal value integral come from neighborhoods around the simple poles $\zeta=\pm q_{0}$ (recall $\lambda(\pm q_{0})=0$). It will be convenient to define 
\begin{equation}
    \mathcal{F}(\zeta)=-\frac{\zeta(\zeta_{1}-q_{0}^{2}\zeta^{-1})^{2}}{16\pi|\zeta-\zeta_{1}|^{4}\big(k(\zeta)-k_{1}\big)}\langle\Upsilon(\zeta),W\rangle,
\end{equation}
and split the integral into two parts 
\begin{equation}
\label{foc_split}
    \tilde{q}^{+}\sim \mathcal{F}(-q_{0})\underbrace{\dashint_{-\infty}^{0}\frac{1-e^{4i\lambda(\zeta)(k(\zeta)-k_{1})t}}{\lambda(\zeta)^2}e^{2i\frac{\lambda(\zeta)}{\Lambda_{1}}\xi}d\zeta}_{I_{-}}+\mathcal{F}(q_{0})\underbrace{\dashint_{0}^{\infty}\frac{1-e^{4i\lambda(\zeta)(k(\zeta)-k_{1})t}}{\lambda(\zeta)^2}e^{2i\frac{\lambda(\zeta)}{\Lambda_{1}}\xi}d\zeta}_{I_{+}}.
\end{equation}
After simplification, including using \eqref{e:ZY_q0}, we have 
\begin{equation}
\label{e:cal_F}
    \mathcal{F}(\pm q_{0})=\pm\frac{i\alpha e^{i\theta}}{16\pi(k_{1}\mp q_{0})},
\end{equation}
where $\alpha$ is defined by
\begin{equation}
\label{e:alpha_def}
\left\langle\begin{bmatrix}u^{*}\\u\end{bmatrix},W\right\rangle=:i\alpha,\qquad\alpha\in\mathbb{R}.
\end{equation}
Note that due to the form of $W$, the inner product in \eqref{e:alpha_def} is purely imaginary regardless of the choice of perturbation. 
The integrals $I_\pm$ in \eqref{foc_split} are computed explicitly through lengthy but straightforward calculations detailed in Appendix~\ref{SC}. This calculation also naturally reveals the right boundary of the shelf to be $\xi\sim2\Lambda_{1}(k_{1}+q_{0})t$. In particular, putting everything together into \eqref{foc_split}, we find that $\tilde{q}^{+}\sim0$ for $\xi\gg2\Lambda_{1}(k_{1}+q_{0})t$, and inside the shelf region $1\ll\xi\ll2\Lambda_{1}(k_{1}+q_{0})t$ we have
\begin{equation}
    \tilde{q}^{+}\sim\frac{\alpha e^{i\theta}}{8q_{0}(k_{1}+q_{0})}+\frac{\alpha e^{i\theta}}{4\Lambda_{1}(k_{1}+q_{0})}i\xi+\frac{\alpha e^{i\theta}}{2}it.
\end{equation}
This suggests that there is inevitable growth in $\xi$ and $t$ in the first order correction resulting from the singularities at $\pm q_{0}$.  Now, according to our original ansatz \eqref{e:approx}, the full approximate solution in the right shelf region should be expressed as
\begin{equation}
\label{e:approx_2}
    q^{+}\sim e^{i\sigma_{1}}\left[q_{0}e^{i\theta}+\varepsilon\tilde{q}^{+}\right].
\end{equation}
Substituting the expression for $\tilde{q}^{+}$, \eqref{e:approx_2} is equivalent up to $\mathcal{O}(\varepsilon)$ to
\begin{equation}
    q^{+}\sim e^{i\theta+i\sigma_{1}}\left[q_{0}+\varepsilon\frac{\alpha}{8q_{0}(k_{1}+q_{0})}\right]\left[1+\frac{\alpha}{4q_{0}\Lambda_{1}(k_{1}+q_{0})}i\varepsilon\xi+\frac{\alpha}{2q_{{0}}}i\varepsilon t\right].
\end{equation}
We can account for the terms that grow with $T=\varepsilon t$ and $\varepsilon\xi$ by moving them into the exponent, i.e., by approximating the above as
\begin{equation}
    q^{+}\sim \left[q_{0}+\frac{\alpha}{8q_{0}(k_{1}+q_{0})}\right]\exp\left\{i\theta+i\sigma_{1}+\frac{\alpha}{4q_{0}\Lambda_{1}(k_{1}+q_{0})}i\varepsilon\xi+\frac{\alpha}{2q_{{0}}}i\varepsilon t\right\}.
\end{equation}
The purpose of the introduction of the $T$-dependent phase parameter $\sigma_{1}$ is now evident. Namely, we can choose the slow time evolution of the soliton phase to be
\begin{equation}
   \label{e:sigma_T_2}
    \sigma_{1T}=-\frac{\alpha}{2q_{0}},
\end{equation}
such that the secular growth in time resulting from the inevitable time dependence of the first order correction integral is suppressed.
Therefore, we construct the full approximate solution in the right shelf region as
\begin{equation}
    q^{+}\sim (q_{0}+\varepsilon h^{+})e^{i\theta+i\varepsilon\varphi^+},
\end{equation}
where the height of the shelf and the spatial phase gradient are respectively given by
\begin{equation}
    h^{+}=\frac{\alpha}{8q_{0}(k_{1}+q_{0})},\qquad\varphi_{\xi}^{+}=\frac{\alpha}{4q_{0}\Lambda_{1}(k_{1}+q_{0})}.
\end{equation}
We later empirically demonstrate for several examples that this approximation of the spatial and temporal growth of the first order correction as a complex exponential does accurately reflect the dynamics, at least for a moderate time interval. That said, we hypothesize that the breakdown of this approximation could be a reason why our predictions for the soliton phase, which coincide with the ones in \cite{FrantzPRSA2011}, do not necessarily retain their validity up to $\mathcal{O}(1/\varepsilon)$, as we will discuss in more details in Sec~\ref{S7} and in the conclusions. 

Note that \eqref{e:sigma_T_2} serves as a generic formula for the evolution of the soliton phase parameter, which depends (through $\alpha$, see \eqref{e:alpha_def}) on the form of the perturbation. Recall that the orthogonality conditions from Sec~\ref{S5} provide three differential equations for four parameters. When supplemented with this formula, the soliton parameters are completely determined.
Also, observe from the above that the formation of the shelf is directly connected to the evolution of $\sigma_{1}$. Namely, if for a given perturbation we have $\alpha=0$, then no shelf develops and $\sigma_{1}$ is constant (see for example the self-steepening perturbation discussed in Sec~\ref{S7.4}).

Repeating these calculations on the left side of the shelf, i.e., in the asymptotic limit $\xi\rightarrow-\infty$, we find that the left boundary of the shelf is $\xi\sim2\Lambda_{1}(k_{1}-q_{0})t$, with $\tilde{q}^{-}\sim0$ for $\xi\ll2\Lambda_{1}(k_{1}-q_{0})t$, and in the left shelf region $2\Lambda_{1}(k_{1}-q_{0})t\ll\xi\ll-1$:
\begin{equation}
        \tilde{q}^{-}\sim-\frac{\alpha e^{-i\theta}}{8q_{0}(k_{1}-q_{0})}+\frac{\alpha e^{-i\theta}}{4\Lambda_{1}(k_{1}-q_{0})}i\xi+\frac{\alpha e^{-i\theta}}{2}it.
\end{equation}
Similarly, we construct the full approximate solution in the shelf region as
\begin{equation}
    q^{-}\sim (q_{0}+\varepsilon h^{-})e^{i\theta+i\varepsilon\varphi^-},
\end{equation}
where
\begin{equation}
    h^{-}=-\frac{\alpha}{8q_{0}(k_{1}-q_{0})},\qquad \varphi_{\xi}^{-}=\frac{\alpha}{4q_{0}\Lambda_{1}(k_{1}-q_{0})}.
\end{equation}
From these formulae it can be seen that, regardless of the perturbation, we have
\begin{equation}
    \Lambda_{1}\varphi_{\xi}^{\pm}=\pm 2h^{\pm},
\end{equation}
which is in agreement with the results in \cite{FrantzPRSA2011}.

\section{Applications}
\label{S7}
In this section, we apply our results to the perturbed NLS equation \eqref{e:NLS_pert} with various small forcing perturbations: linear damping, nonlinear damping (also referred to as two-photon absorption, or TPA, in nonlinear optics, which is the terminology used in \cite{FrantzPRSA2011}), dissipation, and self-steepening. All the results are corroborated by direct numerical simulations, and compared with the results of the direct perturbation theory, and with the earlier works using perturbation theory based on the squared eigenfunctions.

Throughout the following subsections, we compare our predictions with numerical simulations of the perturbed defocusing NLS equation on a nonzero background. The numerical computations were performed using the method proposed by Kassam and Trefethen in \cite{KassamTrefethen}, which is a modification of the fourth-order exponential time differencing (ETDRK4) method (see also \cite{CoxMatthews,Trefethen}) that employs contour integration to avoid numerical instabilities characteristic of stiff PDEs. The spatial discretization is done using a Fourier integral representation. As such, the (non-decaying) initial condition must be multiplied by a rapidly decaying bump function to justify the use of the Fourier basis. To ensure that any disturbances resulting from this truncation remain sufficiently far from the dark soliton, we use a large computational domain with the length scale of the bump function being orders of magnitude larger than the length scale of the perturbed dark soliton and its shelf. For a typical run, the width of the bump function was taken to be on the order of $10^{4}$. The time step and grid spacing were both taken to be $\sim10^{-3}$.

\subsection{Linear damping}
\label{S7.1}

\begin{figure}[ht!]
\centering
\includegraphics[width=.85\textwidth]{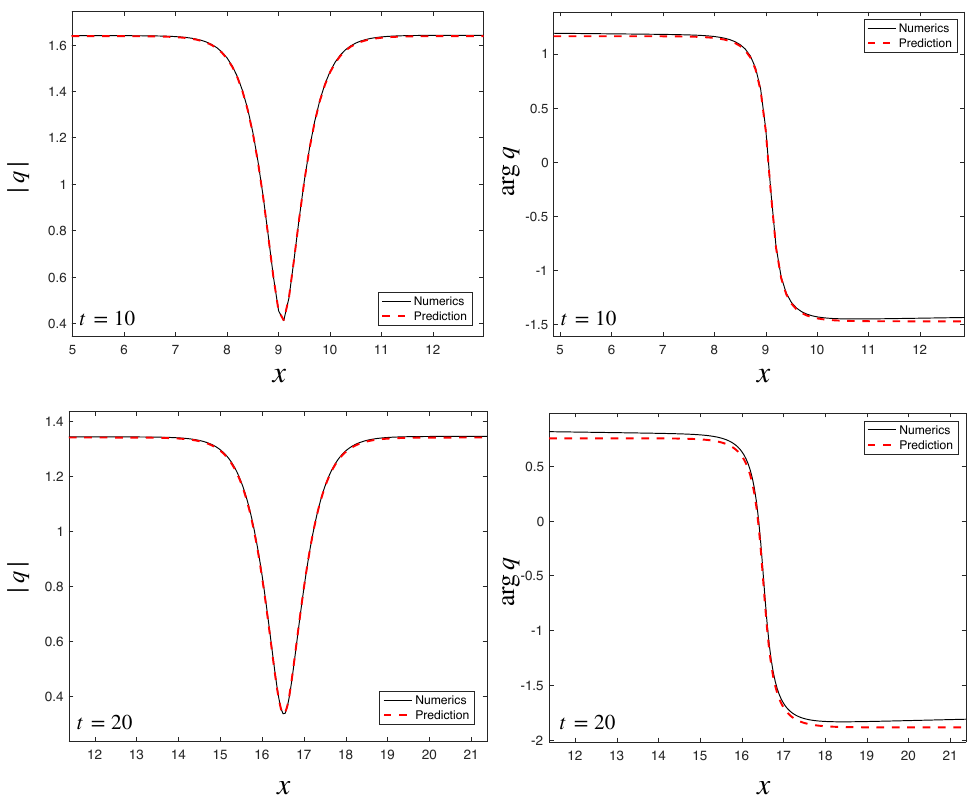}
\caption{Top Row: Comparison of the modulus (left) and phase (right) of the predicted (dashed red) and numerical (solid black) solutions for a dark soliton under the influence of the linear damping perturbation $F[q]=-iq$ with $\varepsilon=0.02$ at time $t=10$. The initial soliton parameters are $q_{0}(0)=2$, $k_{1}(0)=-1/2$, $x_{1}(0)=\sigma_{1}(0)=0$. Bottom Row: The same at $t=20$, in which a slight deviation in our prediction for the phase from the numerics is already visible.}
\label{f:LD_Core}
\end{figure}

Consider Eq.~\eqref{e:NLS_pert} with a linear damping perturbation:
\begin{equation}
    F_{0}=-iu=-ik_{1}+\Lambda_{1}\tanh\xi.
\end{equation}
In this case, \eqref{e:k_ev} implies that
\begin{equation}
   k_{1T}=-k_{1}.
\end{equation}
We now demonstrate that two distinct constraints on the soliton parameters arise from the second orthogonality condition. To this end, a straightforward calculation shows that
\begin{equation}
   \langle\Upsilon_{1}',W\rangle =-\frac{4q_{0}}{\zeta_{1}}\int_{-\infty}^{\infty}\Big\{\left[\Im(e^{-i\theta}F_{0})-\Re(e^{-i\theta}D)\right]e^{\xi}\sech\xi  
   +\frac{k_{1}}{q_{0}}\left[\Im(F_{0})-\Re(D)\right]\xi\sech^{2}\xi\Big\}d\xi,
        \label{e:ip_F0}
\end{equation}
 where $D=e^{-i\sigma_{1}}\partial_{T}(ue^{i\sigma_{1}})$, which is given explicitly in \eqref{e:u_T}.
Note that
\bse
\label{e:DF}
\begin{eqnarray}
    \Re(D)&=&k_{1T}-\sigma_{1T}\Lambda_{1}\tanh\xi,\\
    \Re(e^{-i\theta}D)&=&\frac{1}{q_{0}}\Big(k_{1}k_{1T}-\sigma_{1T}\Lambda_{1}k_{1}(\tanh\xi-1) \\
    &&\qquad+\,\Lambda_{1}\Lambda_{1T}(\tanh\xi+\xi\sech^{2}\xi)-x_{1T}\Lambda_{1}^{3}\sech^{2}\xi\Big),
    \nonumber
\end{eqnarray}
where we have used $\cos\theta=k_{1}/q_{0}$ and $\sin\theta=\Lambda_{1}/q_{0}$, and in the linear damping case we have
\begin{equation}
    \Im(F_{0})=-k_{1},\qquad
    \Im(e^{-i\theta}F_{0})=-\frac{1}{q_{0}}\Big(k_{1}^{2}+\Lambda_{1}^{2}\tanh\xi\Big).
\end{equation}
\ese
Substituting \eqref{e:DF} into \eqref{e:ip_F0} and computing all convergent integrals gives
\begin{equation}
\label{e:orth_half_1}
    \langle{\Upsilon}'_{1},W\rangle = 4\Lambda_{1}\zeta_{1}^{-1}\left\{(\Lambda_{1T}+\Lambda_{1})\int_{-\infty}^{\infty}e^{\xi}\tanh\xi\sech\xi \,d\xi+\sigma_{1T}k_{1}+\Lambda_{1T}-2x_{1T}\Lambda_{1}^{2}\right\}.
\end{equation}
The integral that remains is divergent.
Next, we find
\begin{equation}
    \langle(u^{*},u)^{T},W\rangle
    =-\frac{2i}{\Lambda_{1}}\int_{-\infty}^{\infty}\Big\{|u|^{2}+\Re(u^{*}D)\Big\}d\xi,
\end{equation}
with
\bse
\begin{gather}
|u|^{2}=k_{1}^{2}+\Lambda_{1}^{2}\tanh^{2}\xi,\\
    \Re(u^{*}D)=k_{1}k_{1T}+\Lambda_{1}\Lambda_{1T}\tanh\xi(\tanh\xi+\xi\sech^{2}\xi)-x_{1T}\Lambda_{1}^{3}\tanh\xi\sech^{2}\xi.
\end{gather}
\ese
Substituting these and computing all convergent integrals gives
\begin{equation}
    \label{e:orth_half_2}
    \langle(u^{*},u)^{T},W\rangle=-2i\left\{(\Lambda_{1T}+\Lambda_{1})\int_{-\infty}^{\infty}\tanh^{2}\xi \,d\xi+\Lambda_{1T}\right\}.
\end{equation}
Again, the remaining integral is divergent.
To eliminate the divergence from both inner products above, it suffices to enforce: 
\begin{equation}
 \Lambda_{1T}=-\Lambda_{1}.
\end{equation}
With $\Lambda_{1}$ now fully determined, \eqref{e:orth_half_1} and \eqref{e:orth_half_2} reduce to 
\bse
\begin{gather}
   \label{e:orth_half_3}
    \langle{\Upsilon}'_{1},W\rangle=4\Lambda_{1}\zeta_{1}^{-1}(\sigma_{1T}k_{1}-\Lambda_{1}-2x_{1T}\Lambda_{1}^{2}),\\
       \label{e:orth_half_4}
       \langle(u^{*},u)^{T},W\rangle=2i\Lambda_{1}.
\end{gather}
\ese
We note that the same linear damping perturbation was studied in \cite{ChenJPA1997}. In that work, there were two important differences in the approach compared to ours: (i) the possibility of slow evolution of the phase independent of the other soliton parameters, which we account for with $\sigma_{1T}$, was not considered; (ii) the discrete squared eigenfunctions were not properly modified to account for the contributions from $\pm q_{0}$, that is, the orthogonality condition being enforced was $\langle\Upsilon'_{1},W\rangle=0$. For these reasons, the conclusion in \cite{ChenJPA1997} is that the evolution of the soliton center is given by $x_{1T}=-1/2\Lambda_{1}$. This prediction is not consistent with physical intuition (or numerical simulations), as it is independent of the direction that the soliton is traveling.
Putting \eqref{e:orth_half_3} and \eqref{e:orth_half_4} together as in \eqref{e:up_tilde}, we find the proper orthogonality condition to be
\begin{equation}
    \langle\tilde\Upsilon_{1},W\rangle=4\Lambda_{1}\zeta_{1}^{-1}(\sigma_{1T}k_{1}-2x_{1T}\Lambda_{1}^{2})=0,
\end{equation}
which provides another condition on the soliton parameters, namely,
\begin{equation}
    \sigma_{1T}k_{1}=2x_{1T}\Lambda_{1}^{2}.
\end{equation}
Finally, upon supplementing this condition with the formula for $\sigma_{1T}$ obtained in \eqref{e:sigma_T_2}, all four soliton parameters are uniquely determined. In particular, using \eqref{e:orth_half_4}, we find
\begin{equation}
    \label{e:phase_damp}
    \sigma_{1T}=-\frac{\Lambda_{1}}{q_{0}}.
\end{equation}
The corresponding evolution of the soliton center is
\begin{equation}
    \label{e:center_damp}
x_{1T}=-\frac{k_{1}}{2q_{0}\Lambda_{1}}.
\end{equation}
Solving for the explicit time evolution of all soliton parameters:
\bse
\label{e:evolution_all_damping}
\begin{eqnarray}
    &q_{0}(T)=q_{0}(0)e^{-T},\qquad
    k_{1}(T)=k_{1}(0)e^{-T},\qquad 
    \Lambda_{1}(T)=\Lambda_{1}(0)e^{-T},\\
    &\displaystyle\sigma_{1}(T)=-\frac{\Lambda_{1}(0)}{q_{0}(0)}T+\sigma_{1}(0),\qquad 
    x_{1}(T)=\frac{k_{1}(0)}{2q_{0}(0)\Lambda_{1}(0)}(1-e^{T})+x_{1}(0).
\end{eqnarray}
\ese
All of these formulae agree exactly with those found using a different method in \cite{FrantzPRSA2011}, after the appropriate notational changes, with the exception of the soliton center, which was determined in \cite{FrantzPRSA2011} by employing information from $\mathcal{O}(\varepsilon^{2})$.
Figure~\ref{f:LD_Core} shows a comparison of the modulus and phase of the predicted solution with a numerical simulation for $\varepsilon=0.02$ with snapshots taken at moderate times. 

\begin{figure}[ht!]
\centering
\includegraphics[width=.85\textwidth]{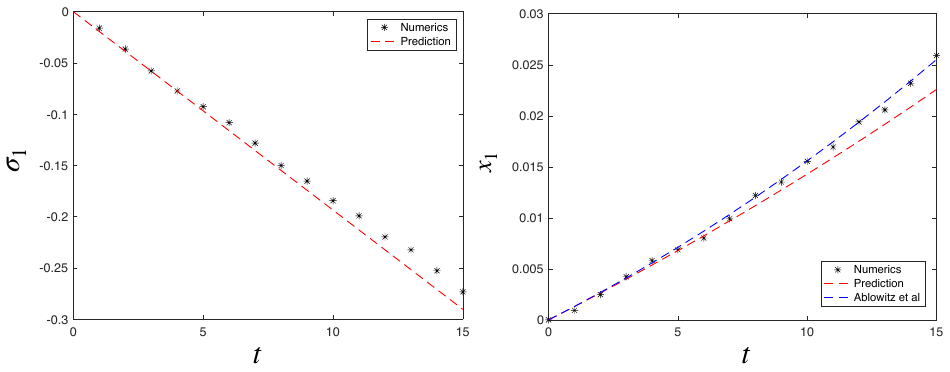}
\caption{The predicted evolution of the parameters $\sigma_{1}$ (left) and $x_{1}$ (right) with time (dashed red), compared with numerical measurements (dotted black). The parameters are the same as in Figure~\ref{f:LD_Core}. Note that both predictions are accurate for a moderate time interval, but start to deviate as time increases. The blue dashed line in the right panel shows the prediction of the method of \cite{FrantzPRSA2011} for the soliton center, which is seen to be more accurate for long time. A similar deviation in the prediction for the soliton phase for larger times can be seen in Fig.~7 in \cite{FrantzPRSA2011}.}
\label{f:LD_Parameters}
\end{figure}

Note that here and in all subsequent examples, since the background evolves on the slow time scale, for the sake of comparison we numerically solve the version of the NLS equation without the background factored out (i.e., \eqref{e:NLS} without the $q_{0}^{2}$ term) and express the approximate solution as
\begin{equation}
\label{e:prediction}
    q(x,t)\approx\left\{k_{1}(T)+i\Lambda_{1}(T)\tanh\left[\Lambda_{1}(T)\left(x+2\int_{0}^{t}k_{1}\,ds-x_{1}(T)\right)\right]\right\}e^{i\sigma_{1}(T)-2i\int_{0}^{t}q_{0}^{2}\,ds}.
\end{equation}
Figure~\ref{f:LD_Parameters} shows our predicted evolution of the soliton center and phase with $t$, compared to numerical measurements of the corresponding quantities. 

From this figure it is clear that, as mentioned in the introduction, the predictions for the center and phase begin to deviate from the numerics before $t=\mathcal{O}(1/\varepsilon)$. We postulate that this breakdown in the approximation could be due to the secular growth in the first order correction integral, and that one would likely be required to compute the $\mathcal{O}(\varepsilon^{2})$ term in the expansion to extend the time interval of validity. Note that the prediction for the soliton center presented in \cite{FrantzPRSA2011}, which is given in terms of a second-order differential equation found using information at $\mathcal{O}(\varepsilon^{2})$, does remain accurate for longer time intervals (though the same discrepancy in the phase is present in that work). It is worth noting that in Figure~\ref{f:LD_Parameters}, and in some subsequent figures, one can perceive a slight staggering of the points corresponding to the numerical measurements of the center and phase. This is due to the fact that the center could lie between two grid points (the spacing of which was constrained in practice due to the necessity for a very large computational domain), and is not an error in the simulation itself. The error is on the order of the grid spacing, and does not affect the overall trend.

\begin{figure}[ht!]
\centering
\includegraphics[width=.85\textwidth]{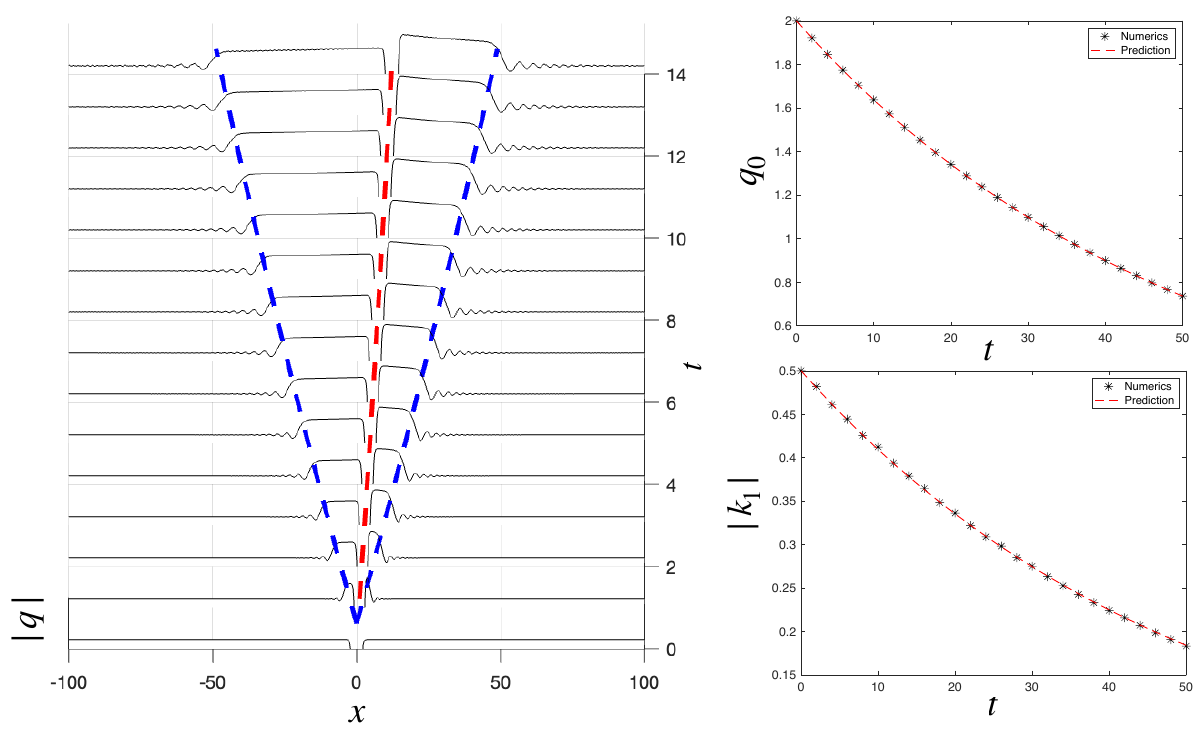}
\caption{A space-time plot showing the development of a raised shelf around the soliton under linear damping. The blue dashed lines denote the boundaries of the shelf region, and the red dashed line corresponds to the predicted path of the soliton core, accounting for the slow evolution of the velocity. For visualization purposes, the evolution of the background has been artificially removed. In the right panel, a comparison of the predicted decrease in the background $q_{0}$ (top) and trough amplitude $|k_{1}|$ (bottom) with numerical measurements are shown up to $t=1/\varepsilon=50$. The initial parameters are the same as in Figure~\ref{f:LD_Core}.}
\label{f:LD_Spacetime}
\end{figure}

\begin{figure}[ht!]
\centering
\includegraphics[width=.85\textwidth]{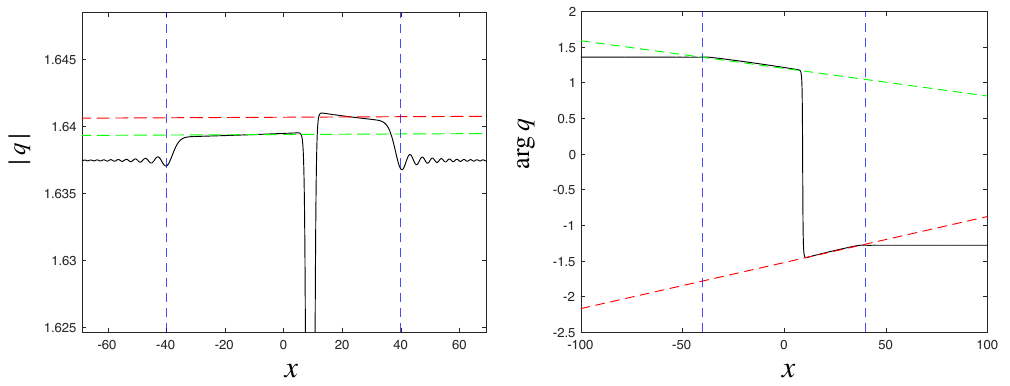}
\caption{The modulus (left) and phase (right) of a numerical simulation of a dark soliton under the influence of the linear damping perturbation $F[q]=-iq$ with $\varepsilon=0.02$ at $t=10$. The blue dashed lines denote the predicted boundaries of the shelf region. The red (resp., green) dashed lines represent the predictions for the shelf height and phase gradient on the right (resp., on the left) of the soliton. The initial parameters are the same as in Figure~\ref{f:LD_Core}.}
\label{f:LD_Shelf}
\end{figure}

Lastly, the predicted height and phase gradient of the shelf in the linear damping case are given by
\begin{equation}
    h^{\pm}=\pm\frac{\Lambda_{1}}{4q_{0}(k_{1}\pm q_{0})},\qquad\varphi_{\xi}^{\pm}=\frac{1}{2q_{0}(k_{1}\pm q_{0})}.
\end{equation}
From these formulae, it can be seen that since $|k_{1}|<q_{0}$ the shelf is raised on both sides of the soliton, and the phase gradient is positive on the right side and negative on the left side of the soliton. In Figure~\ref{f:LD_Spacetime}, we show a space-time plot from which the development of the shelf around the soliton and its propagation can be seen. Figure~\ref{f:LD_Shelf} shows a snapshot of the shelf region, along with the predictions for the height and phase gradient.

\subsection{Nonlinear damping}
\label{S7.2}

Consider Eq.~\eqref{e:NLS_pert} with a nonlinear damping perturbation
\begin{equation}
    F_{0}=-i|u|^{2}u=(q_{0}^{2}-\Lambda_{1}^{2}\sech^{2}\xi)(-ik_{1}+\Lambda_{1}\tanh\xi).
\end{equation}
In this case, \eqref{e:k_ev} implies that
\begin{equation}
k_{1T}=-\left(\frac{2}{3}k_{1}^{2}+\frac{1}{3}q_{0}^{2}\right)k_{1}.
\end{equation}
Following similar steps as in Sec~\ref{S7.1}, we substitute
\begin{equation}
    \Im(e^{-i\theta}F_{0})=-\frac{1}{q_{0}}(q_{0}^{2}-\Lambda_{1}^{2}\sech^{2}\xi)(k_{1}^{2}+\Lambda_{1}^{2}\tanh\xi),
\end{equation}
into \eqref{e:ip_F0}, and after simplification we find
\begin{equation}
\label{e:orth_half_1_non}
    \langle\Upsilon_{1}',W\rangle = \frac{4\Lambda_{1}}{\zeta_{1}}\Bigg\{\left(\Lambda_{1}q_{0}^{2}+\frac{2}{3}\Lambda_{1}k_{1}^{2}+\Lambda_{1T}\right)\int_{-\infty}^{\infty}e^{\xi}\tanh\xi\sech\xi \,d\xi +\Lambda_{1T}-\frac{2}{3}\Lambda_{1}q_{0}^{2}+\sigma_{1T}k_{1}-2x_{1T}\Lambda_{1}^{2}\Bigg\}. 
\end{equation}
Furthermore, one can verify that
\begin{equation}
    \label{e:orth_half_2_non}
    \langle(u^{*},u)^{T},W\rangle=-2i\left\{\left(\Lambda_{1}q_{0}^{2}+\frac{2}{3}\Lambda_{1}k_{1}^{2}+\Lambda_{1T}\right)\int_{-\infty}^{\infty}1d\xi-4q_{0}^{2}\Lambda_{1}+\frac{4}{3}\Lambda_{1}^{3}-\Lambda_{1T}\right\}.
\end{equation}
To eliminate the divergence in both \eqref{e:orth_half_1_non} and \eqref{e:orth_half_2_non}, set
\begin{equation}
\label{e:lambda_ev_nonlinear}
\Lambda_{1T}=-\left(\frac{2}{3}k_{1}^{2}+q_{0}^{2}\right)\Lambda_{1}.
\end{equation}
Note that combining the evolution of $k_{1}$ and $\Lambda_{1}$, we get the evolution of the background
\begin{equation}
    q_{0T}=-q_{0}^{3}.
\end{equation}
 With $\Lambda_{1}$ determined, \eqref{e:orth_half_1_non} and \eqref{e:orth_half_2_non} become
\bse
\begin{gather}
   \label{e:orth_half_3_non}
    \langle\Upsilon_{1}',W\rangle= \frac{4\Lambda_{1}}{\zeta_{1}}\left\{-\left(\frac{2}{3}k_{1}^{2}+\frac{5}{3}q_{0}^{2}\right)\Lambda_{1}+\sigma_{1T}k_{1}-2x_{1T}\Lambda_{1}^{2}\right\},\\
       \label{e:orth_half_4_non}
       \langle(u^{*},u)^{T},W\rangle=2i\left(\frac{2}{3}k_{1}^{2}+\frac{5}{3}q_{0}^{2}\right)\Lambda_{1}.
\end{gather}
\ese
Combining, we find
\begin{equation}
    \langle\tilde\Upsilon_{1},W\rangle=4\Lambda_{1}\zeta_{1}^{-1}(\sigma_{1T}k_{1}-2x_{1T}\Lambda_{1}^{2}),
\end{equation}
which is the same as in the previous example. So, we must again enforce the condition
\begin{equation}
    \sigma_{1T}k_{1}=2x_{1T}\Lambda_{1}^{2}.
\end{equation}
From \eqref{e:sigma_T_2}, using \eqref{e:orth_half_4_non} we find the evolution of the phase to be
\begin{equation}
\label{e:phase_non}
    \sigma_{1T}=-\left(\frac{2}{3}k_{1}^{2}+\frac{5}{3}q_{0}^{2}\right)\frac{\Lambda_{1}}{q_{0}},
\end{equation}
which in turn determines the evolution of the soliton center to be
\begin{equation}
    \label{e:center_non}
x_{1T}=-\left(\frac{2}{3}k_{1}^{2}+\frac{5}{3}q_{0}^{2}\right)\frac{k_{1}}{2q_{0}\Lambda_{1}}.
\end{equation}
The evolution equations for the velocity, amplitude, and phase coincide with the ones obtained from the method of \cite{FrantzPRSA2011} (note that no prediction for the soliton center is given in \cite{FrantzPRSA2011} for this example).

\begin{figure}[ht!]
\centering
\includegraphics[width=.85\textwidth]{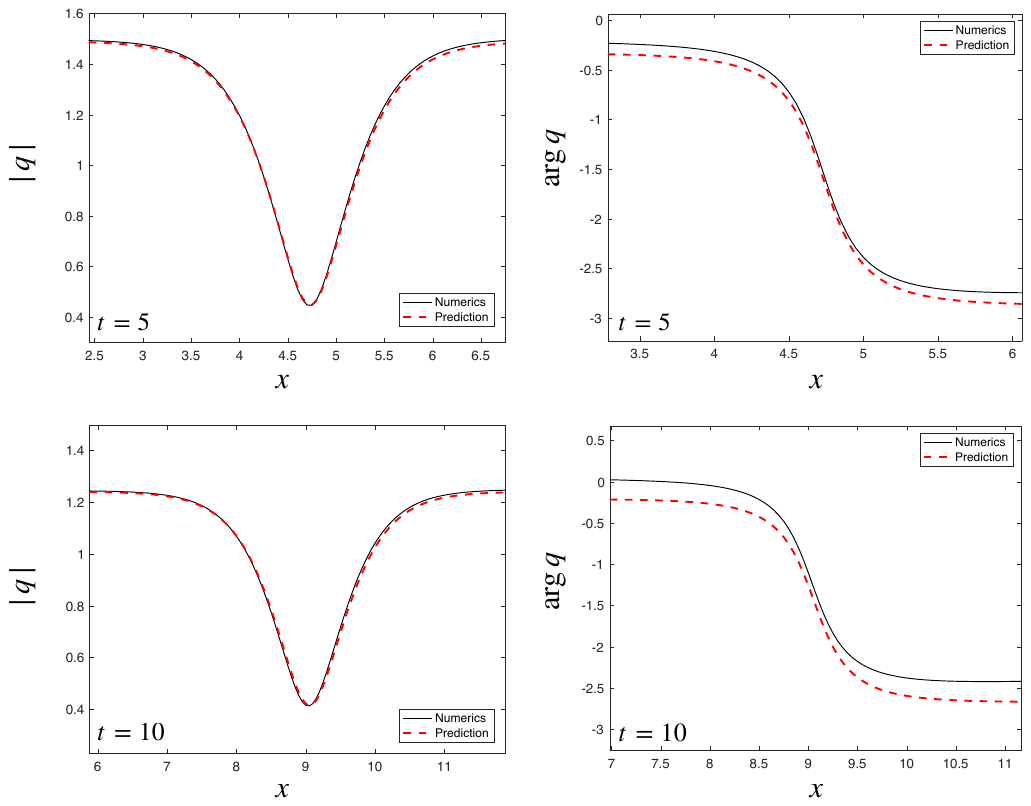}
\caption{Top Row: Comparison of the modulus (left) and phase (right) of the predicted (dashed red) and numerical (solid black) solutions for a dark soliton under the influence of the nonlinear damping perturbation $F[q]=-i|q|^{2}q$ with $\varepsilon=0.02$ at time $t=5$. The initial soliton parameters are $q_{0}(0)=2$, $k_{1}(0)=-1/2$, $x_{1}(0)=\sigma_{1}(0)=0$. Bottom Row: The same at $t=10$, from which a deviation in the phase can be seen.}
\label{f:ND_Core}
\end{figure}
\begin{figure}[H]
\centering
\includegraphics[width=.85\textwidth]{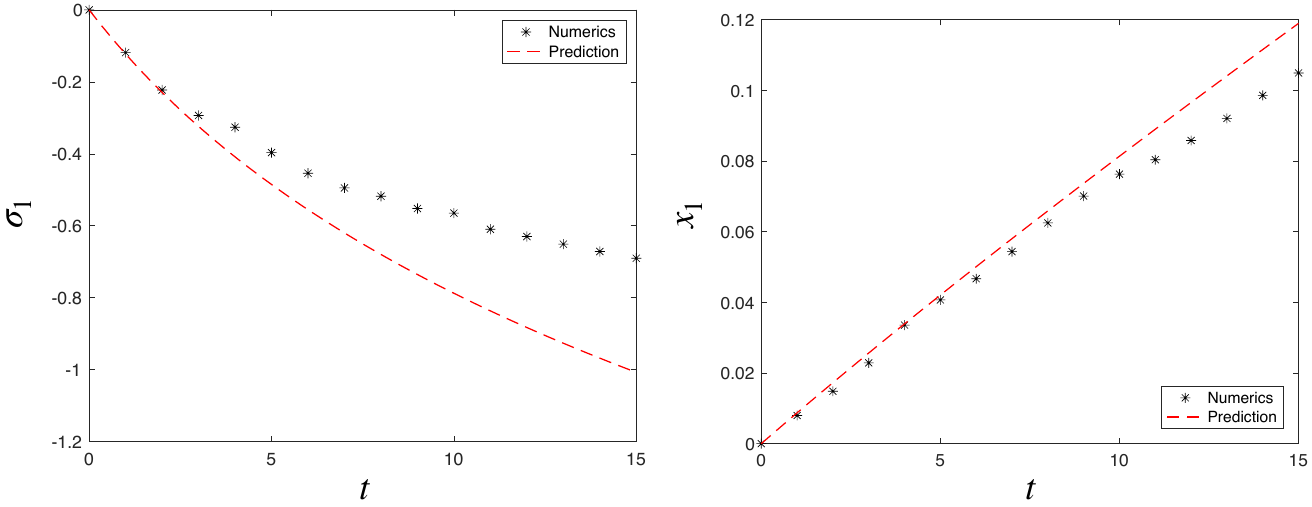}
\caption{The predicted evolution of the parameters $\sigma_{1}$ (left) and $x_{1}$ (right) with time (dashed red), compared with numerical measurements (dotted black). The initial parameters are the same as in Figure~\ref{f:ND_Core}.}
\label{f:ND_Parameters}
\end{figure}
\begin{figure}[ht!]
\centering
\includegraphics[width=.85\textwidth]{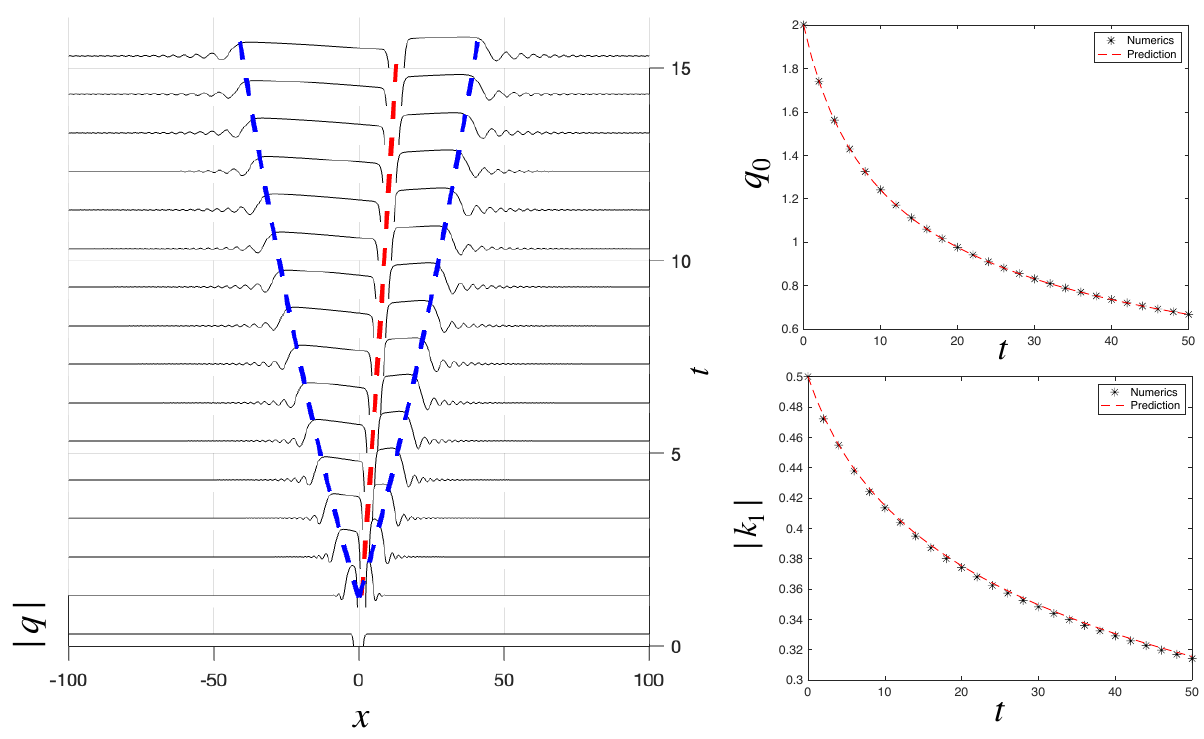}
\caption{A space-time plot showing the development of a raised shelf around the soliton under nonlinear damping. The blue dashed lines denote the boundaries of the shelf region, and the red dashed line corresponds to the predicted path of the soliton core, accounting for the slow evolution of the velocity. For visualization purposes, the evolution of the background has been artificially removed. In the right panel, a comparison of the predicted decrease in the background $q_{0}$ (top) and trough amplitude $|k_{1}|$ (bottom) with numerical measurements are shown up to $t=1/\varepsilon=50$. The initial parameters are the same as in Figure~\ref{f:ND_Core}.}
\label{f:ND_Spacetime}
\end{figure}
\begin{figure}[ht!]
\centering
\includegraphics[width=.85\textwidth]{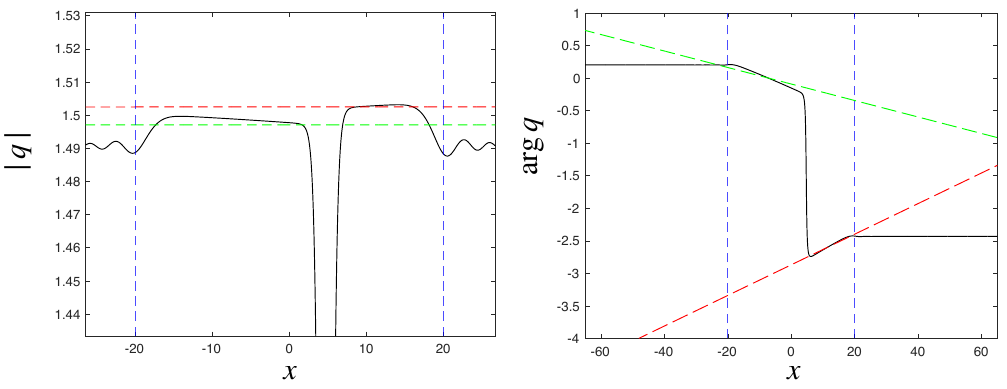}
\caption{The modulus (left) and phase (right) of a numerical simulation of a dark soliton under the influence of the nonlinear damping perturbation $F[q]=-i|q|^{2}q$ with $\varepsilon=0.02$ at $t=5$. The blue dashed lines denote the predicted boundaries of the shelf region. The red (resp., green) dashed lines represent the predictions for the shelf height and phase gradient on the right (resp., on the left) of the soliton. The initial parameters are the same as in Figure~\ref{f:ND_Core}.}
\label{f:ND_Shelf}
\end{figure}

\newpage
In this case, the system of nonlinear differential equations for the soliton parameters must be solved numerically. Figure~\ref{f:ND_Core} shows snapshots of a comparison between our predictions (with the evolution equations for the soliton parameters solved numerically and inserted into \eqref{e:prediction}) with a simulation. Qualitatively, the nonlinear damping case resembles the linear damping case, with the prediction for the phase showing a noticeable discrepancy from the numerics before $t=\mathcal{O}(1/\varepsilon)$. While the prediction for the center holds up reasonably well, a deviation does occur, illustrated in Figure~\ref{f:ND_Parameters}. Finally, Figure~\ref{f:ND_Spacetime} shows a spacetime plot of the development of the shelf, and Figure~\ref{f:ND_Shelf} shows the shelf region and the predictions for the height and phase gradient, which in this case are given by
\begin{equation}
    h^{\pm}=\pm\frac{\Lambda_{1}}{4q_{0}(k_{1}\pm q_{0})}\left(\frac{2}{3}k_{1}^{2}+\frac{5}{3}q_{0}^{2}\right),\qquad\varphi_{\xi}^{\pm}=\frac{1}{2q_{0}(k_{1}\pm q_{0})}\left(\frac{2}{3}k_{1}^{2}+\frac{5}{3}q_{0}^{2}\right).
\end{equation}
Like in the linear damping case, the shelf is raised on both sides of the soliton. 

\subsection{Dissipation}
\label{S7.3}

\begin{figure}[ht!]
\centering
\includegraphics[width=.85\textwidth]{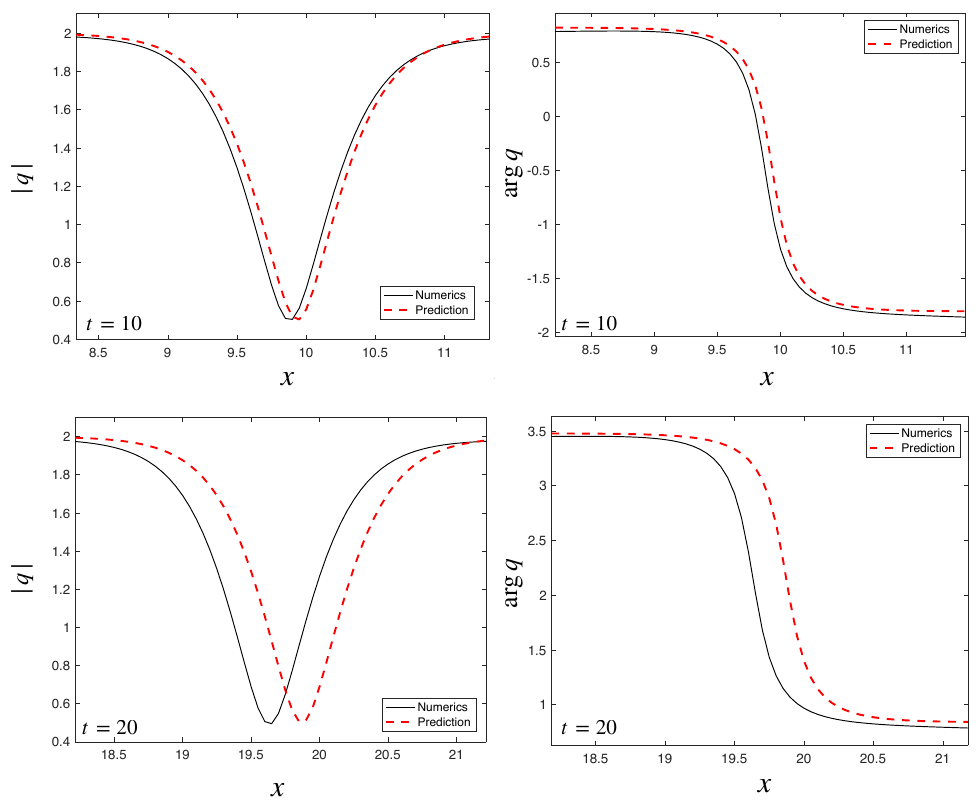}
\caption{Top Row: Comparison of the modulus (left) and phase (right) of the predicted (dashed red) and numerical (solid black) solutions for a dark soliton under the influence of the dissipative perturbation $F[q]=iq_{xx}$ with $\varepsilon=0.02$ at time $t=10$. The initial soliton parameters are $q_{0}(0)=2$, $k_{1}(0)=-1/2$, $x_{1}(0)=\sigma_{1}(0)=0$. Bottom Row: The same at $t=20$, from which it can be seen that our prediction for the soliton center deviates by an amount consistent with a comparison with \cite{FrantzPRSA2011}, which suggests that the next correction to the center would be quadratic in $T$. Note that the shift in the phase in the bottom right panel is due to the shift in the center; aside from this, the phase prediction remains accurate, as can be seen from the left panel in Fig.~\ref{f:DP_Parameters}.}
\label{f:DP_Core}
\end{figure}

Consider Eq.~\eqref{e:NLS_pert} with a dissipative perturbation
\begin{equation}
    F_{0}=iu_{xx}=2\Lambda_{1}^{3}\sech^{2}\xi\tanh\xi.
\end{equation}
In this case, since $F_{0}$ is purely real, \eqref{e:k_ev} implies that
\begin{equation}
  k_{1T}=0.
\end{equation}
Returning to \eqref{e:ip_F0} and putting in the new expression for $F_{0}$ gives
\begin{gather}
    \langle\Upsilon_{1}',W\rangle =\frac{4}{\zeta_{1}}\int_{-\infty}^{\infty}\Big\{2\Lambda_{1}^{4}e^{\xi}\sech^{3}\xi\tanh\xi+\Lambda_{1}\Lambda_{1T}(e^{\xi}\tanh\xi\sech\xi+\xi e^{\xi}\sech^{3}\xi) \\
     -x_{1T}\Lambda_{1}^{3}e^{\xi}\sech^{3}\xi
    -\sigma_{1T}\Lambda_{1}k_{1}(\tanh\xi-1)e^{\xi}\sech\xi-k_{1}\sigma_{1T}\Lambda_{1}\xi\tanh\xi\sech^{2}\xi\Big\}d\xi.\notag
\end{gather}
Computing the convergent integrals, this reduces to
\begin{equation}
\label{e:orth_half_1_dis}
    \langle\Upsilon_{1}',W\rangle = 4\Lambda_{1}\zeta_{1}^{-1}\left\{\Lambda_{1T}\int_{-\infty}^{\infty}e^{\xi}\tanh\xi\sech\xi \,d\xi+\frac{4}{3}\Lambda_{1}^{3}+\sigma_{1T}k_{1}+\Lambda_{1T}-2x_{1T}\Lambda_{1}^{2}\right\}.
\end{equation}
Next, we find 
\begin{equation}
    \label{e:orth_half_2_dis}
    \langle(u^{*},u)^{T},W\rangle=-2i\left\{\Lambda_{1T}\int_{-\infty}^{\infty}\tanh^{2}\xi \,d\xi+\frac{4}{3}\Lambda_{1}^{3}+\Lambda_{1T}\right\}.
\end{equation}
The only way to remove the divergence from both \eqref{e:orth_half_1_dis} and \eqref{e:orth_half_2_dis} is to set
\begin{equation}
    \Lambda_{1T}=0.
\end{equation}
With this, we have
\bse
\begin{gather}
   \label{e:orth_half_3_dis}
    \langle\dot{\Upsilon}(\zeta_{1}),W\rangle= 4\Lambda_{1}\zeta_{1}^{-1}\left(\frac{4}{3}\Lambda_{1}^{3}+\sigma_{1T}k_{1}-2x_{1T}\Lambda_{1}^{2}\right),\\
       \label{e:orth_half_4_dis}
       \langle(u^{*},u)^{T},W\rangle=-\frac{8}{3}i\Lambda_{1}^{3}.
\end{gather}
\ese
Putting these together, we find
\begin{equation}
    \langle\tilde\Upsilon,W\rangle=4\Lambda_{1}\zeta_{1}^{-1}(\sigma_{1T}k_{1}-2x_{1T}\Lambda_{1}^{2}),
\end{equation}
which leads to the same condition on the center and phase that was found in both the previous examples, 
\begin{equation}
    \sigma_{1T}k_{1}=2x_{1T}\Lambda_{1}^{2}.
\end{equation}
Substituting \eqref{e:orth_half_4_dis} into \eqref{e:sigma_T_2} gives
\begin{equation}
    \label{e:phase_dis}
\sigma_{1T}=\frac{4\Lambda_{1}^{3}}{3q_{0}},
\end{equation}
and the corresponding evolution of the soliton center is
\begin{equation}
    \label{e:center_dis}
x_{1T}=\frac{2\Lambda_{1}k_{1}}{3q_{0}}.
\end{equation}
Solving explicitly for all soliton parameters:
\bse
\label{e:evolution_all_dissipation}
\begin{eqnarray}
    &q_{0}(T)=q_{0}(0),\;\;\;k_{1}(T)=k_{1}(0),\;\;\;\Lambda_{1}(T)=\Lambda_{1}(0),\\
    &\displaystyle\sigma_{1}(T)=\frac{4\Lambda_{1}(0)^{3}}{3q_{0}(0)}T+\sigma_{1}(0),\;\;\;x_{1}(T)=\frac{2\Lambda_{1}(0)k_{1}(0)}{3q_{0}(0)}T+x_{1}(0).
\end{eqnarray}
\ese
Figure~\ref{f:DP_Core} shows a comparison of our prediction with a numerical simulation. Note that in this case, the discrepancy in the soliton center at larger times is serious. This is due to the fact that since the background, velocity, and amplitude are constant, our prediction for the evolution of the center is limited to a linear function of $T$. In \cite{FrantzPRSA2011}, a quadratic prediction that is valid for longer time intervals is obtained, but we again note that the method used in that work employs information at $\mathcal{O}(\varepsilon^{2})$ to determine the center, suggesting that a higher order correction would be needed in the present method as well. The reason why our theory does not capture the quadratic term is explained in Remark 3 below.

\begin{figure}[ht!]
\centering
\includegraphics[width=.85\textwidth]{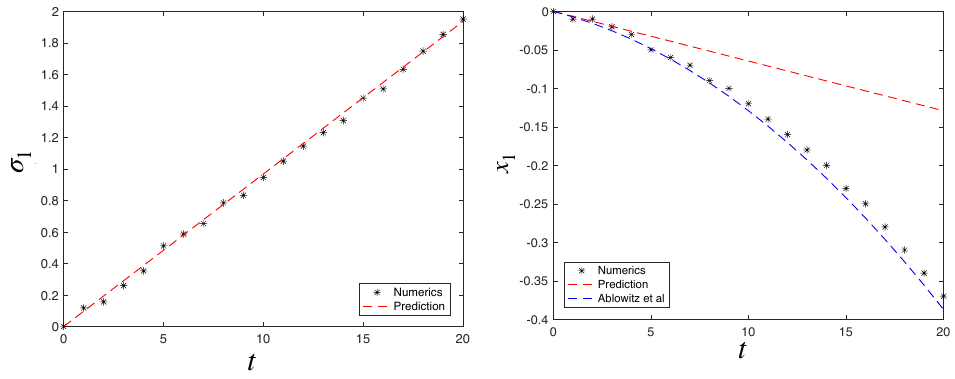}
\caption{The predicted evolution of the parameters $\sigma_{1}$ (left) and $x_{1}$ (right) with time (dashed red), compared with numerical measurements (dotted black) under a dissipative perturbation. The parameters are the same as in Figure~\ref{f:DP_Core}. Note that the prediction for the phase remains accurate for long times, but the linear prediction for the center only remains valid for a short time interval. The blue dashed line in the right panel shows the quadratic prediction for the soliton center obtained using the method of \cite{FrantzPRSA2011}, which accurately captures the curvature.}
\label{f:DP_Parameters}
\end{figure}

The discrepancy is illustrated in Figure~\ref{f:DP_Parameters}, where the predicted phase (which does remain accurate for long times) and center are compared to numerical measurements. Note that our linear prediction for the center captures the initial slope of the curve.

\begin{figure}[ht!]
\centering
\includegraphics[width=.85\textwidth]{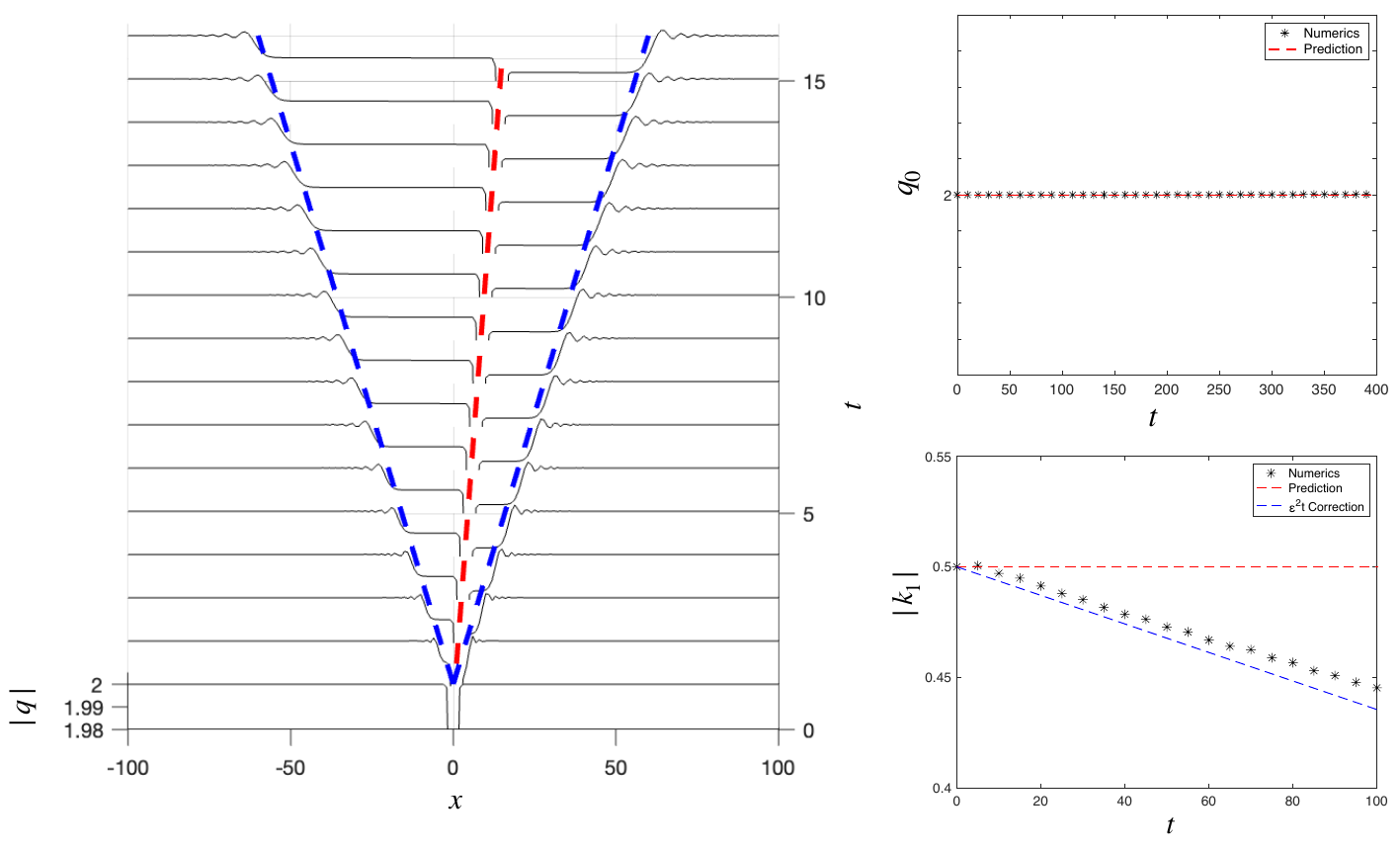}
\caption{A space-time plot showing the development of a depressed shelf around the soliton under a dissipative perturbation. The blue dashed lines denote the boundaries of the shelf region, and the red dashed line corresponds to the path of the soliton core. In the top right panel, the background amplitude $q_{0}$ is shown to remain constant up to at least $t=400$. Each mark on the vertical axis represents an increment of $10^{-9}$. In bottom right panel, the slow decrease in the trough amplitude $|k_{1}|$ is shown up to $t=100$. The numerical measurements are compared to the hypothetical $\varepsilon^{2}t$-dependent formula given in \eqref{e:k1_eps_sq}, which was obtained by interpreting the quadratic correction to the soliton center found in \cite{FrantzPRSA2011} instead as a next-order correction to the velocity, as described in Remark 3. The initial parameters are the same as in Figure~\ref{f:DP_Core}.}
\label{f:DP_Spacetime}
\end{figure}

\begin{figure}[ht!]
\centering
\includegraphics[width=.85\textwidth]{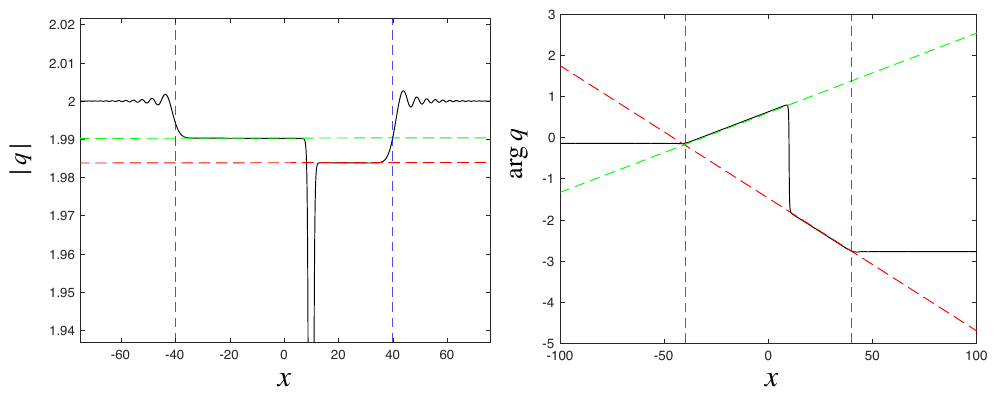}
\caption{The modulus (left) and phase (right) of a numerical simulation of a dark soliton under the influence of the dissipative perturbation $F[q]=iq_{xx}$ with $\varepsilon=0.02$ at $t=10$. The blue dashed lines denote the predicted boundaries of the shelf region. The red (resp., green) dashed lines represent the predictions for the shelf height and phase gradient on the right (resp., on the left) of the soliton. The initial parameters are the same as in Figure~\ref{f:DP_Core}.}
\label{f:DP_Shelf}
\end{figure}

\begin{remark}
    In \cite{FrantzPRSA2011}, $\mathcal{O}(\varepsilon^{2})$ information is used to obtain the following formula (after notational changes) for the soliton center, which is plotted in blue in the right panel of Figure~\ref{f:DP_Parameters}:
    \begin{equation}
    \label{e:center_abl}
        x_{1}(T)=x_{1}(0)+\frac{2\Lambda_{1}(0)k_{1}(0)}{3q_{0}(0)}T+\frac{8\Lambda_{1}(0)^{3}k_{1}(0)}{9q_{0}(0)}T^{2}.
    \end{equation}
    A possible explanation as to why the present theory is unable to capture the quadratic term in \eqref{e:center_abl} is that this term could arise from a dependence of the velocity parameter $k_{1}$ on the slower time scale $\hat T=\varepsilon T=\varepsilon^2 t$. In particular, since
    \begin{equation}
        \frac{8\Lambda_{1}(0)^{3}k_{1}(0)}{9q_{0}(0)}T^{2}=2\int_{0}^{T}\frac{8\Lambda_{1}(0)^{3}k_{1}(0)}{9q_{0}(0)}\, \hat{s}\,d\hat{s},\nonumber
    \end{equation}
    an identical correction to the location of the soliton center can be obtained by taking the evolution of the velocity parameter to be
    \begin{equation}
    \label{e:k1_eps_sq}
        k_{1}(\hat{T})=k_{1}(0)-\frac{8\Lambda_{1}(0)^{3}k_{1}(0)}{9q_{0}(0)}\hat{T}.
    \end{equation}
    Recall that the parameter $k_{1}$ controls not only the velocity but also the trough amplitude of the dark soliton. As such, if the correction found in \cite{FrantzPRSA2011} for the center is to be interpreted instead as a correction to $k_{1}$ as in \eqref{e:k1_eps_sq}, then there should be an observable slow decrease in the trough amplitude over longer time scales. Numerical evidence suggests that this is the case, and that \eqref{e:k1_eps_sq} could provide an appropriate correction to the trough amplitude. In the bottom right panel of Figure~\ref{f:DP_Spacetime}, numerical measurements of the trough amplitude are plotted in comparison with the hypothetical $\varepsilon^2 t$-dependent correction to $k_{1}$.

    By similar logic, on account of \eqref{e:prediction} we also speculate that a discrepancy in the phase parameter $\sigma_{1}$ (e.g., in the case of linear and nonlinear damping) could arise due to $\hat{T}$-dependence of the background amplitude $q_{0}$. In the present case of dissipation, the numerics indicate that $q_{0}$ does in fact remain constant over long time intervals, and no dependence on $\hat{T}$ is detected, as shown in the top right panel of Figure~\ref{f:DP_Spacetime}. This is consistent with the fact that our prediction for $\sigma_{1}$ does remain accurate in this case (cf. Figure~\ref{f:DP_Parameters}).
\end{remark}
The height of the shelf and the phase gradient are given by
\begin{equation}
    h^{\pm}=\mp\frac{\Lambda_{1}^{3}}{3q_{0}(k_{1}\pm q_{0})},\qquad\varphi_{\xi}^{\pm}=-\frac{2\Lambda_{1}^{2}}{3q_{0}(k_{1}\pm q_{0})}.
\end{equation}
In this case, we see that since $|k_{1}|<q_{0}$, the shelf is depressed on both sides of the soliton, and the phase gradient is positive on the left and negative on the right. A space-time plot showing the development and propagation of the shelf is displayed in Figure~\ref{f:DP_Spacetime}, and a snapshot of the shelf region including the predicted height and phase gradient is shown in Figure~\ref{f:DP_Shelf}.

\subsection{Self-steepening}
\label{S7.4}

\begin{figure}[ht!]
\centering
    \includegraphics[width=.85\textwidth]{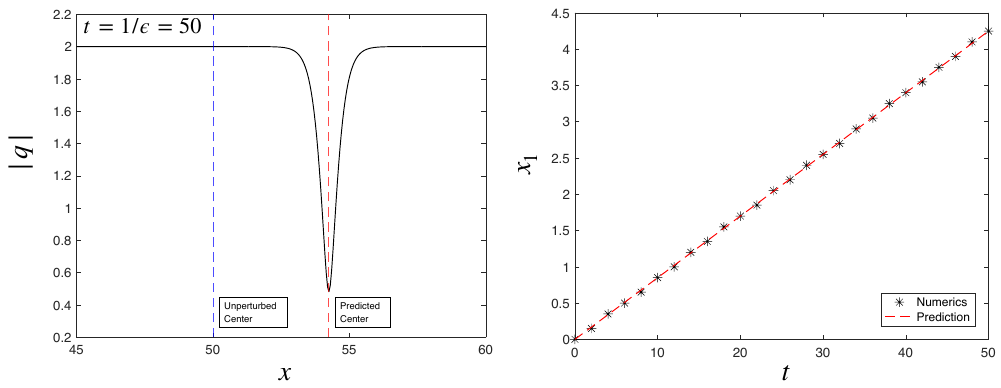}
    \caption{Left: The modulus of a numerical simulation of a dark soliton under the influence of the self-steepening perturbation $F[q]=-i(|q|^{2}q)_{x}$ with $\varepsilon=0.02$ at $t=50$. The red dashed line denotes the location of the predicted soliton center, while the blue dashed line marks where the center would have been in the absence of the perturbation. Right: A comparison of our prediction for the evolution of the center with the numerically measured values. The initial soliton parameters are $q_{0}(0)=2$, $k_{1}(0)=-1/2$, $x_{1}(0)=\sigma_{1}(0)=0$.}
    \label{f:SS_Center}
\end{figure}

We note that while the condition $\sigma_{1T}k_{1}=2x_{1T}\Lambda_{1}^{2}$ has appeared in all three previous examples, it is not universal. For example, in the case of a self-steepening perturbation
\begin{equation}
    F_{0}=-i(|u|^{2}u)_{x}=-\Lambda_{1}^{2}\big(2ik_{1}\Lambda_{1}\sech^{2}\xi\tanh\xi-k_{1}^{2}\sech^{2}\xi-3\Lambda_{1}\sech^{2}\xi\tanh^{2}\xi\big),
\end{equation}
one finds after applying our method that $k_{1T}=0$, $\Lambda_{1T}=0$, $\sigma_{1T}=0$ and no prominent shelf is formed. The absence of the shelf is tied to the fact that for this perturbation, we have that
\begin{equation}
\label{e:pm_orth}
    \langle\Upsilon_{0}^{\pm},W\rangle=0.
\end{equation}
Thus, the singularity in the first order correction integral \eqref{e:correction_3} is fully removed, and in turn there is no secular growth. For the same reason, any perturbation satisfying \eqref{e:pm_orth} will not generate a shelf. 
In the self-steepening case, the only soliton parameter affected is the center, whose evolution is obtained directly from the orthogonality condition $\langle\tilde\Upsilon_{1},W\rangle=0$ and is given by
\begin{equation}
\label{e:steepening_center}
    x_{1T}=2k_{1}^{2}+\Lambda_{1}^2.
\end{equation}
So, the center is pushed forward linearly by the perturbation, which is similar to the situation for bright solitons under a self-steepening effect (see \cite{Yang}). We note that our predictions here are significantly different than those given in \cite{ChenCPL1998} for the same perturbation. In that work, the amplitude and velocity of the dark soliton are predicted to evolve as a result of the perturbation. However, this is not supported by the numerics, which confirm that these parameters remain constant under the perturbation. Figure~\ref{f:SS_Center} demonstrates that our prediction for the soliton center \eqref{e:steepening_center} remains accurate even up to $t=1/\varepsilon$.

\section{Conclusions}
\label{S8}
In this work, we revisited the integrable perturbation theory of the scalar defocusing NLS on a nontrivial background, addressing the conjecture posed in \cite{FrantzPRSA2011} that, because of the existence of an expanding shelf that develops on each side of the soliton due to the perturbation, the squared eigenfunctions associated with the soliton are an insufficient basis, and questioning the existence of a closure relation for this problem, and the ability of integrable perturbation theory to accurately describe the radiation shelf.

We provided a proof of the completeness of the squared eigenfunctions, a crucial difference with respect to the earlier works lying in properly accounting for the singularities of the scattering data at the points $\pm q_0$, which are branch points for the continuous spectrum.
Moreover, we showed that the first order correction can in fact be used to describe the soliton shelf, which results from the singularities of the scattering data at the points $\pm q_0$, and we obtained estimates for the magnitude, velocity and phase gradient on each side of the shelf. 
Using the 1-soliton closure relation, suitable suppression of secular growth, and the asymptotic behavior of the first order correction we determined the adiabatic evolution of the soliton parameters as functions of a slow time variable $T=\varepsilon t$.

All the results were corroborated by direct numerical simulations, and compared with the results of the direct perturbation theory in \cite{FrantzPRSA2011}, and with the earlier works using perturbation theory based on the squared eigenfunctions. In particular, we showed that the estimates we obtained for all soliton parameters agree with the ones in \cite{FrantzPRSA2011} to order $\varepsilon$, the only difference being for the soliton center, which, unlike the other parameters, obtained from perturbed conserved quantities up to $\mathcal{O}(\varepsilon)$,  is determined in \cite{FrantzPRSA2011} in terms of differential equations obtained from the Hamiltonian at $\mathcal{O}(\varepsilon^2)$. We also explained the shortcomings of the predictions for the adiabatic evolution of the soliton parameters in the earlier works within the framework of the integrable perturbation theory.


For some perturbations, our predictions for the soliton center and phase do not match the numerical data on the time scale of $T=\varepsilon t=\mathcal{O}(1)$. A similar problem for the phase can also be seen in some of the examples considered in \cite{FrantzPRSA2011}.
The reason for this discrepancy has been discussed in Remark~3, and it lies in the fact that since the slow-time-dependent velocity must be integrated in time when passing to the co-moving frame (cf Eq.~\eqref{e:new_xi}), it is actually the case the dependence of $k_1$ on an even slower time scale $\varepsilon^2t$ could contribute nontrivially to the soliton center. Thus, to ensure a correct prediction for the soliton center, one would need to include another time scale in the perturbation theory.  A similar argument would apply to the soliton phase $\sigma_1$, on account of the integration in time of the slow-varying background amplitude $q_0$ in Eq.~\eqref{e:prediction}.
This indicates that in order to achieve 
correct estimates up to  $t=\mathcal{O}(1/\varepsilon)$ for the soliton center and phase, one would need to compute the second order correction terms. It is important to stress that this is an intrinsic feature of the problem, 
and not of the perturbative approach, as both our estimates, based on the integrable perturbation theory, and the ones in  \cite{FrantzPRSA2011}, based on direct perturbation theory, with multiple scale expansions and perturbed conservation laws, exhibit the same behavior. In fact, one could argue that similar considerations should be applied even in the bright soliton perturbation theory, where the same transformation to the co-moving frame is routinely applied.

We believe the present work may open up a new vein of research on integrable perturbation theory for integrable systems on a nonzero background, and the methodology can also be generalized to discrete and coupled integrable systems.
Another natural follow-up is to explore ways to improve the accuracy of the estimates for the adiabatic evolution of the soliton parameters for times up to $\mathcal{O}(1/\varepsilon)$. This could be done by introducing an additional time scale $\varepsilon^2t$ and a second order correction to the potential, which will require handling quadratic contributions in the first order correction appearing in the orthogonality conditions. Alternatively, one could generalize the approach used to determine the slow-time evolution of the soliton center in \cite{FrantzPRSA2011}, which makes use of the Hamiltonian at $\mathcal{O}(\varepsilon^2)$, and consider also the other perturbed conserved quantities to $\mathcal{O}(\varepsilon^2)$ in order to get additional corrections to the soliton phase, amplitude and velocity.  

\section*{Acknowledgments}
BP and NJO gratefully acknowledge partial support for this work from the NSF, under grant DMS-2406626. NJO would like to thank Dr. Sathyanarayanan Chandramouli for assistance with the numerics. BP acknowledges the Fulbright Foundation in Greece and the Fulbright program, and the Mathematics Department of the University of Ioannina, Greece, for the kind hospitality during the revision of this work. The authors would also like to thank the anonymous reviewers for their comments, which helped improve the paper.

\appendix

\section{General completeness relation}
\label{SA}
\begin{figure}[ht!]
\centering
    \includegraphics[width=.6\textwidth]{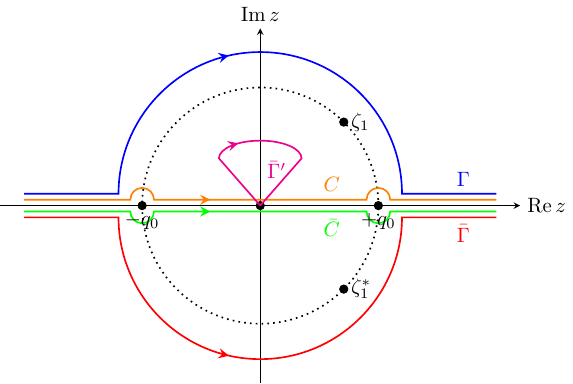}
    \caption{In the upper-half plane, the contour $\Gamma$ (blue) passes above all zeros of $a(z)$ and the contour $C$ (orange) passes below the zeros of $a(z)$ while being indented around the branch points $z=\pm q_{0}$. Their counterparts $\bar\Gamma$ (red) and $\bar{C}$ (green) are shown in the lower-half plane. The contour $\bar\Gamma'$ (magenta) represents the image of $\bar\Gamma$ after the change of variable $z'=q_{0}^{2}/z$.}
    \label{f:Contours}
\end{figure}

To justify the general completeness relation \eqref{e:closure_generic}, note that as $|z|\rightarrow\infty$, we have that $a(z)\sim 1$, $\bar{a}(z)\sim 1$ and
\begin{equation}
    \eta(x,z)\chi(y,z)^T\sim e^{2i\lambda(z)(x-y)}\,\text{diag}\,(0,1),\quad\bar\eta(x,z)\bar\chi(y,z)^T\sim e^{-2i\lambda(z)(x-y)}\,\text{diag}\,(1,0),
\end{equation}
and we recall that $\lambda(z)=(z-q_0^2/z)/2\sim z/2$ and $z\to \infty$.
These are obtained directly from the large-$z$ asymptotic behavior of the Jost eigenfunctions, which can be found, for example, in \cite{P2023,GPT2026} where the same notations as ours are used. From the analyticity properties discussed in Secs~\ref{S2} and \ref{S3} and the above asymptotics, we then get the identities
\bse
\label{e:cont_int}
\begin{eqnarray}
    \int_{\Gamma}\frac{1}{\gamma(\zeta)^{2}a(\zeta)^{2}}\eta(x,\zeta)\chi(y,\zeta)^{T}d\zeta&=&\int_{\Gamma}e^{i\zeta(x-y)}d\zeta\,\text{diag}\,(0,1),\\
    \int_{\bar\Gamma}\frac{1}{\gamma(\zeta)^{2}\bar a(\zeta)^{2}}\bar\eta(x,\zeta)\bar\chi(y,\zeta)^{T}d\zeta&=&\int_{\bar\Gamma}e^{-i\zeta(x-y)}d\zeta\,\text{diag}\,(1,0),
\end{eqnarray}
\ese
where $\Gamma$ is a contour from $-\infty+i0$ to $+\infty+i0$ that passes above all zeros $\zeta_{j}$ of $a(z)$, and $\bar\Gamma$ is a contour from $-\infty-i0$ to $+\infty-i0$ that passes below all zeros $\zeta_{j}^{*}$ of $\bar{a}(z)=a^{*}(z^{*})$. For reference, Figure~\ref{f:Contours} displays the contours $\Gamma$ and $\bar\Gamma$, as well as the additional contours mentioned later in this section. 
Since the integrands on the right-hand sides in \eqref{e:cont_int} are entire functions that decay exponentially as the radius of the contour of integration increases in their respective half-planes, both contours can be deformed to the real axis, and using $\int_{-\infty}^{\infty}e^{\pm i\zeta(x-y)}d\zeta=2\pi\delta(x-y)$ we get
\begin{equation}
    \int_{\Gamma}\frac{1}{\gamma(\zeta)^{2}a(\zeta)^{2}}\eta(x,\zeta)\chi(y,\zeta)^{T}d\zeta+
    \int_{\bar\Gamma}\frac{1}{\gamma(\zeta)^{2}\bar a(\zeta)^{2}}\bar\eta(x,\zeta)\bar\chi(y,\zeta)^{T}d\zeta=2\pi\delta(x-y)I.
\end{equation}
Using \eqref{e:eta_sym_q0} and the similar symmetry satisfied by the adjoint squared eigenfunctions
\begin{equation}
    \label{e:chi_sym_q0}
    \bar\chi(x,z)=\frac{q_{0}^{2}}{z^{2}}e^{-2i\theta}\chi(x,q_{0}^{2}/z),
\end{equation}
as well as $\bar{a}(z)=e^{-2i\theta}a(q_{0}^{2}/z)$, we rewrite the second integral to get
\begin{eqnarray}
\label{e:cont_int_2}
    2\pi\delta(x-y)I&=&\int_{\Gamma}\frac{1}{\gamma(\zeta)^{2}a(\zeta)^{2}}\eta(x,\zeta)\chi(y,\zeta)^{T}d\zeta\nonumber\\&&-\,
    \int_{\bar\Gamma}\frac{1}{\gamma(q_{0}^{2}/\zeta)^{2} a(q_{0}^{2}/\zeta)^{2}}\eta(x,q_{0}^{2}/\zeta)\chi(y,q_{0}^{2}/\zeta)^{T}d\zeta.
\end{eqnarray}
Now, make the change of variable $\zeta'=q_{0}^{2}/\zeta$ in the second integral. The contour $\bar\Gamma$ is mapped to a contour $\bar\Gamma'$ in the upper-half plane that begins at $-0+i0$, travels clockwise while remaining within the circle of radius $q_{0}$, and finally approaches $+0+i0$ (shown in Figure~\ref{f:Contours}). Since all potential poles are on the circle $|z|=q_{0}$, integrating along $\bar\Gamma'$ is equivalent to integrating along $\Gamma$ with the opposite orientation. With this, the two integrals in \eqref{e:cont_int_2} can be combined into 
\begin{equation}
    \int_{\Gamma}\frac{1}{\gamma(\zeta)^{2}a(\zeta)^{2}}\eta(x,\zeta)\chi(y,\zeta)^{T}\left(1-\frac{q_{0}^{2}}{\zeta^{2}}\right)d\zeta=\int_{\Gamma}\frac{1}{\gamma(\zeta)a(\zeta)^{2}}\eta(x,\zeta)\chi(y,\zeta)^{T}d\zeta.
\end{equation}
To explicitly include the contributions from the zeros of $a(z)$, which are double poles of the integrand above, note that
\begin{equation}
    \left(\int_{C}-\int_{\Gamma}\right)\frac{1}{\gamma(\zeta)a(\zeta)^{2}}\eta(x,\zeta)\chi(y,\zeta)^{T}d\zeta=2\pi i\sum_{j=1}^{J}\Res_{z=\zeta_j}\frac{\eta(x,z)\chi(y,z)^{T}}{\gamma(z)a(z)^{2}},
\end{equation}
where $C$ is a contour from $-\infty+i0$ to $+\infty+i0$ that follows the real axis while being indented in the upper half plane to avoid $z=\pm q_{0}$. 
The residue at each $\zeta_{j}$ is found to be:
\begin{gather}
    \frac{1}{\gamma(\zeta_{j})a'(\zeta_{j})^{2}}\Bigg[\frac{\gamma(\zeta_{j})a''(\zeta_{j})-\gamma'(\zeta_{j}){a}'(\zeta_{j})}{\gamma(\zeta_{j}){a}'(\zeta_{j})}\eta(x,\zeta_{j})\chi(y,\zeta_{j})^{T} 
     \\
    +\,\eta'(x,\zeta_{j})\chi(y,\zeta_{j})^{T}+\eta(x,\zeta_{j})\chi'(y,\zeta_{j})^{T}\Bigg]. \nonumber
\end{gather}
Substituting the above expression into \eqref{e:closure_generic} leads to \eqref{e:closure_poles}.

\section{Derivation of 1-soliton Jost eigenfunctions}
\label{SB}
In the case of no reflection and a single soliton ($\rho\equiv0$ and $J=1$), \eqref{e:lin_sys} reduces to the linear algebraic system
\bse
\begin{gather}
\label{e:psi_sys_1}
    \psi(x,t,z)e^{-i\Omega(x,t,z)}=\begin{bmatrix}
        -iq_{+}/z\\1
    \end{bmatrix}+\frac{e^{-i\Omega(x,t,\zeta_{1}^{*})}\bar\psi(x,t,\zeta_{1}^{*})C_{1}^{*}}{z-\zeta_{1}^{*}},\\
\label{e:psi_sys_2}
    \bar\psi(x,t,z)e^{i\Omega(x,t,z)}=\begin{bmatrix}
        1\\iq_{+}^{*}/z
    \end{bmatrix}+\frac{e^{i\Omega(x,t,\zeta_{1})}\psi(x,t,\zeta_{1})C_{1}}{z-\zeta_{1}}.
\end{gather}
\ese
Let the discrete eigenvalue be $\zeta_{1}=k_{1}+i\Lambda_{1}$ with $k_{1}\in\mathbb{R}$, $\Lambda_{1}>0$. 
We need to review here some properties of the scattering coefficient $a(z)$. Taking into account its analyticity properties, its zeros, and the symmetries, one can obtain the following representation (trace formula) for $a(z)$ for $z\in \Complex^+$:
\begin{gather}
\label{e:trace}
a(z)=\prod_{j=1}^J\frac{z-\zeta_j}{z-\zeta_j^*}\exp{\left[ -\frac{1}{2\pi i} \int_{-\infty}^{\infty} \frac{\log(1-\rho(z)\rho^*(\zeta^*))}{\zeta -z}d\zeta \right]}\,,
\end{gather}
where $J$ is the number of (simple) discrete eigenvalues.
Recalling that $a(z)\rightarrow q_+/q_-$ as $z\rightarrow 0$, we conclude that the potential satisfies
\begin{gather}
\frac{q_+}{q_-}=\prod_{j=1}^J\frac{\zeta_j}{\zeta_j^*} \, \exp{\left[-\frac{1}{2\pi i} \int_{-\infty}^{\infty} \frac{\log(1-\rho(z)\rho^*(\zeta^*))}{z}d\zeta \right]}\,,
\end{gather}
which is often referred to as the $\theta$-condition \cite{FT1987}. 
Due to our choice of boundary conditions $q_{\pm}=q_{0}e^{\pm i\theta}$, the  $\theta$-condition with $J=1$ and $\rho(z)\equiv 0$ demands that $\zeta_{1}=q_{+}$. Furthermore, note that $\Omega_{1}:=\Omega(x,t,\zeta_{1})=-\Omega(x,t,\zeta_{1}^{*})$. With this in mind, evaluating \eqref{e:psi_sys_1} at $z=\zeta_{1}$ and \eqref{e:psi_sys_2} at $z=\zeta_{1}^{*}$ gives
\bse
\begin{gather}
\label{e:psi_sys_3}
    \psi(x,t,\zeta_{1})=\begin{bmatrix}
        -i\\1
    \end{bmatrix}e^{i\Omega_{1}}+\frac{e^{2i\Omega_{1}}\bar\psi(x,t,\zeta_{1}^{*})C_{1}^{*}}{\zeta_{1}-\zeta_{1}^{*}},\\
\label{e:psi_sys_4}
    \bar\psi(x,t,\zeta_{1}^{*})=\begin{bmatrix}
        1\\i
    \end{bmatrix}e^{i\Omega_{1}}+\frac{e^{2i\Omega_{1}}\psi(x,t,\zeta_{1})C_{1}}{\zeta_{1}^{*}-\zeta_{1}}.
\end{gather}
\ese
Using $\zeta_{1}-\zeta_{1}^{*}=2i\Lambda_{1}$ and taking $C_{1}\in\mathbb{R}$ according to \eqref{e:2ndsymmCj} with $\alpha_{1}=\theta$, solving for $\bar\psi(x,t,\zeta_{1}^*)$ yields
\begin{equation}
    \bar\psi(x,t,\zeta_{1}^*)=e^{i\Omega_{1}}\begin{bmatrix}
        1\\i
    \end{bmatrix}\frac{2\Lambda_{1}}{2\Lambda_{1}-C_{1}e^{2i\Omega_{1}}}.
\end{equation}
Defining a new parameter (the soliton center) $x_{1}$ through $C_{1}=-2\Lambda_{1}e^{2\Lambda_{1}x_{1}}$, this becomes 
\begin{equation}
    \bar\psi(x,t,\zeta_{1}^*)=e^{i\Omega_{1}}\begin{bmatrix}
        1\\i
    \end{bmatrix}\frac{1}{1+e^{2i\Omega_{1}+2\Lambda_{1}x_{1}}}.
\end{equation}
This can be written in terms of the traveling coordinate defined in \eqref{e:1_sol},
\begin{equation}
\label{e:xi_omega}
    \xi=\Lambda_{1}(x+2k_{1}t-x_{1})=-i\Omega_{1}-\Lambda_{1}x_{1},
\end{equation}
in which case we have
\begin{equation}
\label{e:psi_sech}
    \bar\psi(x,t,\zeta_{1}^*)=e^{i\Omega_{1}}\begin{bmatrix}
        1\\i
    \end{bmatrix}\frac{1}{1+e^{-2\xi}}=e^{i\Omega_{1}}\begin{bmatrix}
        1\\i
    \end{bmatrix}\frac{1}{2}e^{\xi}\sech\xi.
\end{equation}
Substituting \eqref{e:psi_sech} into \eqref{e:psi_sys_1} and again using \eqref{e:xi_omega}, we obtain
\begin{equation}
    \psi(x,t,z)e^{-i\Omega(x,t,z)}=\begin{bmatrix}
        -iq_{+}/z\\1
    \end{bmatrix}-\frac{\Lambda_{1}e^{-2\xi}}{z-\zeta_{1}^{*}}\begin{bmatrix}
        1\\i
    \end{bmatrix}\e^{\xi}\sech\xi.
\end{equation}
By components, this is:
\bse
\begin{gather}
    \psi_{1}(x,t,z)=\Big(-iq_{+}z^{-1}-\frac{1}{z-\zeta_{1}^{*}}\Lambda_{1}e^{-\xi}\sech\xi\Big)e^{i\Omega(x,t,z)},\\
        \psi_{2}(x,t,z)=\Big(1-i\frac{1}{z-\zeta_{1}^{*}}\Lambda_{1}e^{-\xi}\sech\xi\Big)e^{i\Omega(x,t,z)}.
\end{gather}
\ese
The two components of $\bar\psi(x,t,z)$ can then be recovered through the symmetry \eqref{e:phi_psi_sym_1},
\bse
\begin{gather}
    \bar{\psi}_{1}(x,t,z)=\psi_{2}^{*}(x,t,z^{*})=\Big(1+i\frac{1}{z-\zeta_{1}}\Lambda_{1}e^{-\xi}\sech\xi\Big)e^{-i\Omega(x,t,z)},\\
    \bar\psi_{2}(x,t,z)=\psi_{1}^{*}(x,t,z^{*})=\Big(iq_{-}z^{-1}-\frac{1}{z-\zeta_{1}}\Lambda_{1}e^{-\xi}\sech\xi\Big)e^{-i\Omega(x,t,z)}.
\end{gather}
\ese
Finally, the two components of $\phi(x,t,z)$ can be found through the relation \eqref{e:phi_psi},
\bse
\begin{gather}
        \phi_{1}(x,t,z)=\Big(\frac{z-\zeta_{1}}{z-\zeta_{1}^{*}}+i\frac{1}{z-\zeta_{1}^{*}}\Lambda_{1}e^{-\xi}\sech\xi\Big)e^{-i\Omega(x,t,z)},\\
    \phi_{2}(x,t,z)=\Big(iq_{-}z^{-1}\frac{z-\zeta_{1}}{z-\zeta_{1}^{*}}-\frac{1}{z-\zeta_{1}^{*}}\Lambda_{1}e^{-\xi}\sech\xi\Big)e^{-i\Omega(x,t,z)},
\end{gather}
\ese
and the two components of $\bar\phi(x,t,z)$ can be recovered through symmetries,
\bse
\begin{gather}
    \bar\phi_{1}(x,t,z)=\phi_{2}^{*}(x,t,z^{*})=\Big(-iq_{+}z^{-1}\frac{z-\zeta_{1}^{*}}{z-\zeta_{1}}-\frac{1}{z-\zeta_{1}}\Lambda_{1}e^{-\xi}\sech\xi\Big)e^{i\Omega(x,t,z)},\\
    \bar\phi_{2}(x,t,z)=\phi_{1}^{*}(x,t,z^{*})=\Big(\frac{z-\zeta_{1}^{*}}{z-\zeta_{1}}-i\frac{1}{z-\zeta_{1}}\Lambda_{1}e^{-\xi}\sech\xi\Big)e^{i\Omega(x,t,z)}.
\end{gather}
\ese

\section{Evaluation of the integrals in Eq.~\eqref{foc_split}}
\label{SC}
In this Appendix we provide the detailed calculations of the integrals $I_\pm$ in Eq.~\eqref{foc_split}. First, consider the integral $I_{-}$ given by
\begin{equation}
    I_{-}=\dashint_{-\infty}^{0}\frac{1-e^{4i\lambda(\zeta)(k(\zeta)-k_{1})t}}{\lambda(\zeta)^2}e^{2i\frac{\lambda(\zeta)}{\Lambda_{1}}\xi}d\zeta,
\end{equation}
which has a pole at $\zeta=-q_{0}$. First, using $\lambda(\zeta)=\frac{1}{2}(1-q_{0}\zeta^{-1})(\zeta+q_{0})$, this can be written as
\begin{equation}
    I_{-}=4\,\dashint_{-\infty}^{0}\frac{1}{(\zeta+q_{0})^{2}}\frac{1-e^{4i\lambda(\zeta)(k(\zeta)-k_{1})t}}{(1-q_{0}\zeta^{-1})^{2}}e^{2i\frac{\lambda(\zeta)}{\Lambda_{1}}\xi}d\zeta.
\end{equation}
Applying integration by parts,
\begin{equation}
\label{e:ibp}
    I_{-}=4\,\dashint_{-\infty}^{0}\frac{1}{\zeta+q_{0}}\partial_{\zeta}\left\{\frac{1-e^{4i\lambda(\zeta)(k(\zeta)-k_{1})t}}{(1-q_{0}\zeta^{-1})^{2}}e^{2i\frac{\lambda(\zeta)}{\Lambda_{1}}\xi}\right\}d\zeta.
\end{equation}
Expanding the derivative, we can split \eqref{e:ibp} into 
\begin{equation}
    I_{-}=J_{1}+i\xi J_{2}+it J_{3},
\end{equation}
where the three integrals are given by
\bse
\begin{gather}
    J_{1}=\dashint_{-\infty}^{0}\frac{1}{\zeta+q_{0}}\frac{-8q_{0}\zeta^{-2}}{(1-q_{0}\zeta^{-1})^{3}}\big[1-e^{4i\lambda(\zeta)(k(\zeta)-k_{1})t}\big]e^{2i\frac{\lambda(\zeta)}{\Lambda_{1}}\xi}d\zeta,\\
    J_{2}=\dashint_{-\infty}^{0}\frac{1}{\zeta+q_{0}}\frac{1}{(1-q_{0}\zeta^{-1})^{2}}\frac{8\lambda'(\zeta)}{\Lambda_{1}}\big[1-e^{4i\lambda(\zeta)(k(\zeta)-k_{1})t}\big]e^{2i\frac{\lambda(\zeta)}{\Lambda_{1}}\xi}d\zeta,\\
    J_{3}=\dashint_{-\infty}^{0}\frac{1}{\zeta+q_{0}}(-16)\frac{\lambda'(\zeta)(k(\zeta)-k_{1})+\lambda(\zeta){k}'(\zeta)}{(1-q_{0}\zeta^{-1})^{2}}e^{4i\lambda(\zeta)(k(\zeta)-k_{1})t}e^{2i\frac{\lambda(\zeta)}{\Lambda_{1}}\xi}d\zeta.
\end{gather}
\ese
Beginning with $J_{1}$, using the fact that the dominant contribution to the integral comes from the pole $\zeta\approx-q_{0}$, in which case $\lambda(\zeta)\approx\zeta+q_{0}$ and $k(\zeta)\approx-q_{0}$, we get
\begin{equation}
    J_{1}\sim\frac{1}{q_{0}}\,\dashint_{-\infty}^{0}\frac{1}{\zeta+q_{0}}\big[1-e^{-4i(\zeta+q_{0})(k_{1}+q_{0})t}\big]e^{\frac{2i}{\Lambda_{1}}(\zeta+q_{0})\xi}d\zeta
\end{equation}
To evaluate principal value integrals of this form, we can use the formula
\begin{equation}
\label{e:pv_formula}
    \dashint_{-\infty}^{\infty}\frac{e^{ip\vartheta}}{p}dp= i\pi\sgn \vartheta.
\end{equation}
To this end, $J_{1}$ can be split again into
\begin{equation}
    J_{1}\sim\frac{1}{q_{0}}\,\dashint_{-\infty}^{0}\frac{1}{\zeta+q_{0}}e^{\frac{2i}{\Lambda_{1}}(\zeta+q_{0})\xi}d\zeta-\frac{1}{q_{0}}\,\dashint_{-\infty}^{0}\frac{1}{\zeta+q_{0}}e^{\frac{2i}{\Lambda_{1}}(\zeta+q_{0})\left[\xi-2\Lambda_{1}(k_{1}+q_{0})t\right]}d\zeta.
\end{equation}
Applying the formula \eqref{e:pv_formula} gives
\begin{equation}
    J_{1}\sim\frac{\pi i}{q_{0}}\left\{1-\sgn\big[\xi-2\Lambda_{1}(k_{1}+q_{0})t\big]\right\}.
\end{equation}
If $\xi\gg2\Lambda_{1}(k_{1}+q_{0})t$, then $J_{1}\sim0$. In the right shelf region $1\ll\xi\ll2\Lambda_{1}(k_{1}+q_{0})t$, $J_{1}\sim2\pi i/q_{0}$. Following similar steps for $J_{2}$, we obtain
\begin{equation}
    J_{2}\sim\frac{2\pi i}{\Lambda_{1}}\left\{1-\sgn\big[\xi-2\Lambda_{1}(k_{1}+q_{0})t\big]\right\}.
\end{equation}
which gives in the shelf region $J_{2}=4\pi i/\Lambda_{1}$. Lastly, noting that $\lambda'(\zeta)\approx1$ and $k'(\zeta)\approx0$ near the pole, we get
\begin{equation}
    J_{3}\sim4\pi i(k_{1}+q_{0})\,\sgn\big[\xi-2\Lambda_{1}(k_{1}+q_{0})t\big],
\end{equation}
which shows that 
\begin{equation}
    J_{3}\sim\begin{cases}
        4\pi i(k_{1}+q_{0}),\quad&\xi\gg2\Lambda_{1}(k_{1}+q_{0})t\\
        -4\pi i(k_{1}+q_{0}),\quad&1\ll\xi\ll2\Lambda_{1}(k_{1}+q_{0})t
    \end{cases}.
\end{equation}
We now move on to evaluate the integral $I_{+}$, given by
\begin{equation}
    I_{+}=4\,\dashint_{0}^{\infty}\frac{1}{(\zeta-q_{0})^{2}}\frac{1-e^{4i\lambda(\zeta)(k(\zeta)-k_{1})t}}{(1+q_{0}\zeta^{-1})^{2}}e^{2i\frac{\lambda(\zeta)}{\Lambda_{1}}\xi}d\zeta,
\end{equation}
which has a pole at $\zeta=q_{0}$.
Similarly to the previous calculations, applying integration by parts splits the integral into 
\begin{equation}
    I_{+}=H_{1}+i\xi H_{2}+itH_{3},
\end{equation}
where
\bse
\begin{gather}
    H_{1}=\dashint_{0}^{\infty}\frac{1}{\zeta-q_{0}}\frac{8q_{0}\zeta^{-2}}{(1+q_{0}\zeta^{-1})^{3}}\big[1-e^{4i\lambda(\zeta)(k(\zeta)-k_{1})t}\big]e^{2i\frac{\lambda(\zeta)}{\Lambda_{1}}\xi}d\zeta\\
    H_{2}=\dashint_{0}^{\infty}\frac{1}{\zeta-q_{0}}\frac{1}{(1+q_{0}\zeta^{-1})^{2}}\frac{8\lambda'(\zeta)}{\Lambda_{1}}\big[1-e^{4i\lambda(\zeta)(k(\zeta)-k_{1})t}\big]e^{2i\frac{\lambda(\zeta)}{\Lambda_{1}}\xi}d\zeta\\
    H_{3}=\dashint_{0}^{\infty}\frac{1}{\zeta-q_{0}}(-16)\frac{\lambda'(\zeta)(k(\zeta)-k_{1})+\lambda(\zeta){k}'(\zeta)}{(1+q_{0}\zeta^{-1})^{2}}e^{4i\lambda(\zeta)(k(\zeta)-k_{1})t}e^{2i\frac{\lambda(\zeta)}{\Lambda_{1}}\xi}d\zeta.
\end{gather}
\ese
Using the formula \eqref{e:pv_formula}, we find that first two integrals are
\begin{gather}
    H_{1}\sim\frac{\pi i}{q_{0}}\left\{1-\sgn\big[\xi-2\Lambda_{1}(k_{1}-q_{0})t\big]\right\},\\
    H_{2}\sim\frac{2\pi i}{\Lambda_{1}}\left\{1-\sgn\big[\xi-2\Lambda_{1}(k_{1}-q_{0})t\big]\right\}.
\end{gather}
Since we are studying the large positive $\xi$ limit, and $k_{1}<q_{0}$, we have that $H_{1}\sim0$ and $H_{2}\sim0$. Thus, the contributions to the height and phase gradient of the right side of the shelf only come from the pole $\zeta=-q_{0}$. The third integral is found to be
\begin{equation}
    H_{3}\sim4\pi i(k_{1}-q_{0})\sgn\big[\xi-2\Lambda_{1}(k_{1}-q_{0})t\big],
\end{equation}
which is limit gives $H_{3}\sim4\pi i(k_{1}-q_{0})$. To summarize, putting everything into \eqref{foc_split}, we have that if $\xi\gg2\Lambda_{1}(k_{1}+q_{0})t$,
\begin{equation}
    \tilde{q}^{+}\sim-\mathcal{F}(-q_{0})4\pi (k_{1}+q_{0})t-\mathcal{F}(q_{0})4\pi (k_{1}-q_{0})t=0,
\end{equation}
as expected, and in the right shelf region $1\ll\xi\ll2\Lambda_{1}(k_{1}+q_{0})t$,
\begin{equation}
    q^{+}\sim\mathcal{F}(-q_{0})\left[\frac{2\pi i}{q_{0}}-\frac{4\pi}{\Lambda_{1}}\xi+4\pi (k_{1}+q_{0})t\right]-\mathcal{F}(q_{0})4\pi (k_{1}-q_{0})t.
\end{equation}
Putting in \eqref{e:cal_F}, this reduces to the expression given in Sec~\ref{S6}. The calculation of the left side of the shelf from the large negative $\xi$ limit proceeds in an analogous way. In that case, the contributions to the height and phase gradient of the shelf come only from the pole $\zeta=q_{0}$.

\end{document}